\documentclass[11pt]{article}

\usepackage{fullpage}
\usepackage[colorlinks]{hyperref}

\usepackage[leqno]{amsmath}
\usepackage{amsrefs} 
\usepackage{amssymb}

\usepackage{amsthm}
\renewcommand{\le}{\leq}
\renewcommand{\ge}{\geq}

\usepackage{mathtools}
\usepackage{mathrsfs}
\usepackage{dsfont}
\usepackage{bbm}
\usepackage{soul}
\usepackage{bm}
\usepackage{./shortcuts}
\usepackage{./ben}
\usepackage{enumerate}

\usepackage{comment}

\renewcommand{\kclq}{\mathsf{Clique}_{k,n}}

\usepackage[dvipsnames]{xcolor}
\definecolor{corlinks}{RGB}{0,0,150}
\definecolor{cormenu}{RGB}{0,0,150}
\definecolor{corurl}{RGB}{0,0,150}
\definecolor{colortoc}{rgb}{0.65, 0.16, 0.16}

\begin{document}

\newtheorem{theorem}{Theorem}[section]
\newtheorem{proposition}[theorem]{Proposition}
\newtheorem{lemma}[theorem]{Lemma}
\newtheorem{corollary}[theorem]{Corollary}
\newtheorem{conjecture}[theorem]{Conjecture}
\newtheorem{claim}[theorem]{Claim}
\newtheorem{fact}[theorem]{Fact}

\newtheorem{definition}[theorem]{Definition}

\newtheorem{conv}[theorem]{Convention}
\newtheorem{proviso}[theorem]{Proviso}
\newtheorem{remark}[theorem]{Remark}
\newtheorem{observation}[theorem]{Observation}

\title{%
    Monotone Circuit Lower Bounds from 
    Robust Sunflowers
}

\author{%
    Bruno Pasqualotto Cavalar
    \footnote{Email: \texttt{Bruno.Pasqualotto-Cavalar@warwick.ac.uk}}
    \vspace{0.2cm}
    \\{\small University of Warwick\vspace{0.3cm}}
    \and
    Mrinal Kumar
    \footnote{Email: \texttt{mrinalkumar08@gmail.com}}
    \vspace{0.2cm}
    \\{\small IIT Bombay\vspace{0.3cm}}
    \and
    Benjamin Rossman
    \footnote{Email: \texttt{benjamin.rossman@duke.edu}}
    \vspace{0.2cm}
    \\{\small Duke University\vspace{0.3cm}}
}

\maketitle

\begin{abstract}
Robust sunflowers are a generalization of combinatorial sunflowers that
have applications in monotone circuit complexity \cite{rossman_kclq}, DNF
sparsification \cite{gopalan2013dnf}, randomness extractors
\cite{li2018sunflowers}, and recent advances on the Erd\H{o}s-Rado
sunflower conjecture
\cite{alweiss2019improved,lovett2019fromdnf,rao2020}.  The recent
breakthrough of Alweiss, Lovett, Wu and Zhang \cite{alweiss2019improved}
gives an improved bound on the maximum size of a $w$-set system that
excludes a robust sunflower.  In this paper, we use this result to obtain
an $\exp(n^{1/2-o(1)})$ lower bound on the monotone circuit size of an
explicit $n$-variate monotone function, improving the previous best known
$\exp(n^{1/3-o(1)})$ due to Andreev \cite{andreev1987method} and Harnik and Raz \cite{HR00}. We also show an
$\exp(\Omega(n))$ lower bound on the monotone {\em arithmetic} circuit
size of a related polynomial via a very simple proof.  Finally, we introduce a notion of robust
clique-sunflowers and use this to prove an $n^{\Omega(k)}$ lower bound on
the monotone circuit size of the CLIQUE function for all $k \le
n^{1/3-o(1)}$, strengthening the bound of Alon and Boppana
\cite{alon_boppana}. 
\end{abstract}

{
  \hypersetup{linkcolor=colortoc}
  \tableofcontents
}

\section{Introduction}
A monotone Boolean circuit is a Boolean circuit with $\mathsf{AND}$
and $\mathsf{OR}$ gates but no negations ($\mathsf{NOT}$ gates). Although a restricted model of computation,
monotone Boolean circuits seem a very natural model to work with when
computing \emph{monotone} Boolean functions, i.e., Boolean functions $f :
\{0,1\}^n \to \{0,1\}$ such that for all pairs of inputs $(a_1, a_2,
\ldots, a_n) , (b_1, b_2, \ldots, b_n)\in \{0,1\}^n$ where $a_i \leq b_i$
for every $i$, we have $f(a_1, a_2, \ldots, a_n) \leq f(b_1, b_2, \ldots,
b_n)$. Many natural and well-studied Boolean functions such as $\mathsf{Clique}$ and $\mathsf{Majority}$ are monotone.

Monotone Boolean circuits have been very well studied in Computational
Complexity over the years, and continue to be one of the few seemingly
largest natural sub-classes of Boolean circuits for which we have
exponential lower bounds. This line of work started with an
influential paper of Razborov \cite{razborov_boolean_85} from 1985 which proved an
$n^{\Omega(k)}$ lower bound on the size of
monotone circuits computing the $\kclq$ function on $n$-vertex graphs for $k
\leq \log n$; this bound is super-polynomial for $k = \log n$.
Prior to Razborov's result,
super-linear lower bounds for monotone circuits were unknown,
with the best bound being a lower
bound of $4n$ due to Tiekenheinrich \cite{Tk84}. 
Further progress in this
line of work included the results of Andreev \cite{andreev_monotone} who
proved an exponential lower bound for another explicit
function. 
Alon and Boppana \cite{alon_boppana}
extended Razborov's result by proving an 
$n^{\Omega(\sqrt k)}$
lower bound 
for $\smash{\kclq}$ for all $k \le n^{2/3 - o(1)}$.
A second paper of Andreev \cite{andreev1987method} from the same time period proved an $2^{\Omega(n^{1/3}/\log n)}$ lower bound for an explicit $n$-variate monotone function.
Using a different technique, Harnik and Raz \cite{HR00} 
proved a lower bound of $2^{\Omega((n/\log n)^{1/3})}$ for a family of
explicit $n$-variate functions defined using a small probability space of random
variables with bounded independence.
However, modulo improvements to the polylog factor in this exponent, the
state of art monotone circuit lower bounds have been stuck at 
$2^{\Omega(n^{1/3 - o(1)})}$ since 1987.\footnote{Stasys Jukna (personal communication) observed
that Andreev's bound \cite{andreev1987method} can be improved to
$2^{\Omega((n/\sqrt{\log n})^{1/3})}$ using the lower bound criterion of
\cite{jukna1999combinatorics}.}
To this day, the question of proving truly exponential lower
bounds for monotone circuits (of the form $2^{\Omega(n)}$) for an explicit $n$-variate function) remains open! 
(Truly exponential lower bounds for monotone {\em formulas} were obtained only recently \cite{pitassi2017strongly}.)

In the present paper, we are able to improve the best known lower bound
for monotone circuits by proving an
$2^{\Omega(n^{1/2}/\log n)}$
lower bound
for an explicit $n$-variate monotone Boolean function (Section~\ref{sec:harnik_raz}).
The function is based on the same construction
first considered by Harnik and Raz,
but our argument employs the approximation method of Razborov with
recent improvements on robust sunflower bounds~\cite{alweiss2019improved,rao2020}.
By applying the same technique with a variant of robust sunflowers that we
call
clique-sunflowers,
we are able to prove an $n^{\Omega(k)}$ lower bound
for the
$\kclq$
function
when $k \leq n^{1/3-o(1)}$,
thus improving the result of Alon and Boppana
when $k$ is in this range 
(Section~\ref{sec:clique}).
Finally, we are able to prove truly exponential lower bounds in the
monotone arithmetic setting to a fairly general family of polynomials,
which shares some similarities to the functions considered by Andreev and Harnik and Raz
(Section~\ref{sec:arithmetic}).

\subsection{Monotone circuit lower bounds and sunflowers}
The original lower bound
for $\kclq$ due 
to Razborov employed a technique
which came to be known as the
\emph{approximation method}.
Given a
monotone circuit $C$ of
``small size'', 
it consists into constructing gate-by-gate, in a bottom-up fashion,
another circuit $\widetilde{C}$
that approximates $C$ on most inputs of interest.
One then exploits the structure of this \emph{approximator circuit}
to prove that it differs from $\kclq$ 
on most inputs of interest, thus implying that no ``small'' circuit can
compute this function.
This technique was leveraged to obtain lower bounds
for a host of other monotone problems~\cite{alon_boppana}.

A crucial step in Razborov's proof involved the sunflower lemma
due to 
Erd\H{o}s and Rado.
    A family
    $\calf$
    of subsets of $[n]$
    is called a
    \emph{sunflower}
    if
    there exists a set $Y$
    such that
    $F_1 \cap F_2 = Y$
    for every $F_1, F_2 \in \calf$.
    The sets 
    of $\calf$
    are called
    \emph{petals}
    and the set $Y = \bigcap \calf$ is called the
    \emph{core}.
    We say that the family $\calf$ is $\ell$-uniform
    if every set in the family has size $\ell$.
\begin{theorem}[Erd\H{o}s and Rado~\cite{erdos_rado_sunflower}]
    \label{thm:erdos_rado}
    Let $\calf$ 
    be a $\ell$-uniform 
    family of subsets of $[n]$.
    If
    $\abs{\calf} > \ell!(r-1)^\ell$,
    then
    $\calf$
    contains a sunflower of $r$ petals.
\end{theorem}
Informally, the sunflower lemma allows one to prove that a monotone
function can be approximated by one with fewer minterms by means of the
``plucking'' procedure: if the function has too many (more than
$\ell!(r-1)^\ell$) minterms of size $\ell$, 
then it contains a sunflower with $r$ petals;
remove all the petals, replacing them with the core.
One can then prove that this procedure does not introduce many errors.

The notion of {\em robust sunflowers} was introduced by the third author
in \cite{rossman_kclq}, to achieve better bounds via the approximation
method on the monotone circuit size of $\kclq$ when the negative
instances are Erd\H{o}s-R\'enyi random graphs $\mb G_{n,p}$ below the
$k$-clique
threshold.\footnote{Robust sunflowers were called {\em quasi-sunflowers}
in \cite{rossman_kclq,gopalan2013dnf,li2018sunflowers,lovett2019fromdnf}
and {\em approximate sunflowers} in \cite{lovett2019dnf}. Following
Alweiss {\em et al} \cite{alweiss2019improved}, we adopt the new name
{\em robust sunflower}.}
A family $\calf \sseq 2^{[n]}$ is called a
$(p, \eps)$-\emph{robust sunflower}
if
\begin{equation*}
    \Pr_{\rndw \sseq_p [n]}
    \left[ 
        \exists F \in \calf:
            F \sseq \rndw \cup Y
    \right]
    >
    1-\eps,
\end{equation*}
where 
$Y := \bigcap \calf$
and $\rndw$ is a $p$-random subset of $[n]$
(i.e., every element of $[n]$ is contained in $\rndw$
independently with probability $p$).

As remarked in~\cite{rossman_kclq},
every $\ell$-uniform sunflower 
of $r$ petals  is a
$(p, e^{-rp^{\ell}})$-robust sunflower. Moreover, as observed
in~\cite{lovett2019fromdnf}, every
$(1/r, 1/r)$-robust sunflower contains a sunflower of $r$ petals.
A corresponding bound for the appearance of 
robust sunflowers in large families
was also proved in~\cite{rossman_kclq}.

\begin{theorem}[\cite{rossman_kclq}]
    \label{thm:approx_sunflower}
    Let 
    $\calf$ be a $\ell$-uniform family
    such that
    $\abs{\calf} \geq \ell!(2\log(1/\eps)/p)^\ell$.
    Then $\calf$ contains a
    $(p,\eps)$-robust sunflower.
\end{theorem}

For many choice of parameters $p$ and $\eps$,
this bound is better than the one by Erd\H{o}s and Rado, thus
leading to better approximation bounds.
In a recent breakthrough, this result was significantly improved 
by
Alweiss, Lovett, Wu and Zhang~\cite{alweiss2019improved}.
Soon afterwards, 
alternative proofs with slightly improved bounds were given by
Rao\footnote{
    Rao's bound is also slightly stronger in the following sense.
    He shows that, if the random set $\rndw$ is chosen
    uniformly at random among all sets of size $\floor{np}$,
    then we also have
    $
    \Pr
    \left[ 
        \exists F \in \calf:
            F \sseq \rndw \cup Y
    \right]
    >
    1-\eps$.
    However, for our purposes, the $p$-biased case will suffice.
}~\cite{rao2020} and Tao~\cite{tao2020}.
A more detailed discussion can be found in a note by
Bell, Suchakree and Warnke~\cite{bcw21}.

\begin{theorem}[\cite{alweiss2019improved,rao2020,tao2020,bcw21}]
    \label{thm:improved_sunflower}
    There exists a constant $B > 0$ such that the following holds
    for all $p, \eps \in (0,1/2]$.
    Let 
    $\calf$ be an $\ell$-uniform family
    such that
    $\abs{\calf} \geq 
    (B \log(\ell/\eps)/p)^\ell$.
    Then $\calf$ contains a
    $(p,\eps)$-robust sunflower.
\end{theorem}

Theorem~\ref{thm:improved_sunflower} can be verified by combining
the basic structure of Rossman's original argument~\cite{rossman_kclq}
with the main technical estimate of Rao~\cite{rao2020}.
Since the proof does not appear explicitly in any of those papers,
for completeness we give a proof on Appendix~\ref{sec:sf_proof}.

\subsection{Preliminaries}

We denote by
$\blt^n_{= m} \sseq \blt^n$
the set of all $n$-bit binary vectors
with Hamming weight exactly $m$.
We extend the logical operators $\lor$ and $\land$ to
binary strings $x, y \in \blt^n$, as follows:
\begin{itemize}
    \item $(x \land y)_i = x_i \land y_i$, for every $i \in [n]$;
    \item $(x \lor y)_i = x_i \lor y_i$, for every $i \in [n]$.
\end{itemize}

We will say that a distribution $\rndx$ with support in $\blt^n$
is \emph{$p$-biased} or \emph{$p$-random} if
the random variables $\rndx_1,\dots,\rndx_n$
are mutually independent and satisfy
$\Pr[\rndx_i = 1] = p$ for all $i$.
If a distribution $\rndU$ has support in 
$2^{[n]}$,
we will say that $\rndU$
is \emph{$p$-biased} or \emph{$p$-random} if
the 
random Boolean string $\rndx$
such that
$\rndx_i = 1 \iff i \in \rndU$
is $p$-biased.
We 
sometimes
write
$\rndU \sseq_p [n]$
to denote that $\rndU$
is a $p$-biased subset of $[n]$.

We consistently write random objects using
boldface symbols (such as $\rndw$, $\mb G_{n,p}$, etc).
Everything that is not written in boldface is not random.
When taking probabilities or expectation,
the underlying distribution is always the one referred to
by the boldface symbol.
For instance, 
when $i \in [n]$ and $\rndW$ is a $p$-biased subset of $[n]$,
the event $\set{i \in \rndW}$
denotes that the \emph{non-random} element $i$
is contained in the \emph{random} set $\rndW$.

For a Boolean function $f$ and a probability distribution $\mb \mu$ on the
inputs on $f$,
we write $f(\mb \mu)$ to denote the random variable which evaluates $f$
on a
random instance of $\mb \mu$.

In what follows, we will mostly ignore ceilings and floors for the sake of
convenience, since these do not make any substantial difference in the
final calculations.

\section{Harnik-Raz function}
\label{sec:harnik_raz}

The strongest lower bound known for monotone circuits computing
an explicit $n$-variate monotone Boolean function 
is
$\exp\big(\Omega\big((n/\log n)^{1/3}\big)\big)$,
and 
it
was
obtained by Harnik
and Raz~\cite{HR00}.
In this section, 
we
will
prove 
a lower bound of
$\exp(\Omega(n^{1/2}/\log n))$
for the same Boolean function 
they considered.
We apply the \emph{method of approximations}~\cite{razborov_boolean_85}
and the new \emph{robust sunflower}
bound~\cite{alweiss2019improved,rao2020}.
We do not expect that a lower bound better than
$\exp(n^{1/2-o(1)})$ can be obtained by 
the approximation method with robust sunflowers.
This 
limitation
is discussed 
with more detail in Section~\ref{sec:hr_discussion}.

We start by giving a high level outline of the proof. 
We define the Harnik-Raz function $\fnhr : \blt^n \to \blt$
and find 
two distributions 
$\rndy$ and $\rndn$
with support in
$\blt^n$
satisfying the following properties:
\begin{itemize}
    \item $\fnhr$ outputs~1 on $\rndy$ with high probability
        (Lemma~\ref{claim:hr_result});
    \item $\fnhr$ outputs~0 on $\rndn$ with high probability
        (Lemma~\ref{claim:negative_test}).
\end{itemize}
Because of these properties, the distribution $\rndy$ is called
the
\defx{positive test distribution},
and $\rndn$ is called
the
\defx{negative test distribution}.
We also define a set of monotone Boolean functions called
\emph{approximators}, and we show that:
\begin{itemize}
    \item every approximator commits many mistakes on either $\rndy$ or
        $\rndn$ with high
        probability (Lemma~\ref{lemma:hr_approx_err});
    \item every Boolean function computed by a ``small''
        monotone circuit agrees with
        an approximator on both 
        $\rndy$ and $\rndn$ with high probability
        (Lemma~\ref{lemma:hr_approx_correct}).
\end{itemize}
Together these suffice for proving that ``small'' circuits cannot
compute $\fnhr$. 
The crucial part where the robust sunflower result comes
into play is in 
the last two items.

\subsection{Notation for this section}

For 
$A \subseteq [n]$, 
let $x_A \in \blt^n$
be the binary vector with
support in $A$.
For a set
$A \subseteq [n]$, 
let
$\clqind{A}$ be the indicator function satisfying
\begin{equation*}
    \clqind{A}(x) = 1 \iff  x_A \leq x.
\end{equation*}

For a monotone Boolean function
$f : \blt^n \to \blt$,
let
$\mintms(f)$
denote the set of minterms of $f$,
and let
$\mintms_\ell(f) 
:= 
\mintms(f) \cap \blt_{=\ell}^n$.
Elements of
$\mintms_\ell(f)$
are called $\ell$-minterms of $f$.

This notation is valid only in Section~\ref{sec:harnik_raz}
and will be slightly tweaked in Section~\ref{sec:clique}
(Lower Bound for $\kclq$)
for the sake of uniformity of exposition.

\subsection{The function}
\label{sec:hr_the_function}
We now describe the construction of the function 
$\fnhr : \blt^n \to \blt$
considered by Harnik and Raz~\cite{HR00}.
First observe that,
for every $n$-bit monotone Boolean function 
$f$,
there exists a family $\cals \sseq 2^{[n]}$
such that
\begin{equation*}
    f(x_1,\dots,x_n) 
    = 
    D_\cals(x_1,\dots,x_n)
    :=
    \bigvee_{S \in \cals}
    \bigwedge_{j \in S}
    x_{j}.
\end{equation*}
Indeed, $\cals$ can be chosen to be 
the family of the 
coordinate-sets of minterms of $f$.
Now, in order to construct the Harnik-Raz function,
we will suppose
$n$ is a prime number
and let $\F_n = \set{0,1,\dots,n-1}$ be the field of $n$ elements.
Moreover, we fix two positive integers $c$ and $k$
with $c < k < n$. 
For a polynomial $P \in \F_n[x]$, we let
$S_P$
be the set of the valuations of $P$
in each element of $\set{1,2,\dots,k}$
(in other words,
$S_P = \set{P(1),\dots,P(k)}$).
Observe that it is not necessarily the case that 
$\card{S_P}=k$,
since it may happen that $P(i)=P(j)$ for some $i,j$ such that $i \neq j$.
Finally,
we consider the family 
$\hrfml$
defined as
\begin{equation*}
    \hrfml
    :=
    \set{S_P 
        : P \in \F_n[x], 
        \text{$P$ has degree at most $c-1$ and}
        \abs{S_P} 
        \geq k/2
    }.
\end{equation*}
We thus define 
$\fnhr$
as
$\fnhr := D_{\hrfml}$.

We now explain the choice of 
$\hrfml$.
First,
the choice for valuations of polynomials with degree at most $c-1$
is explained by 
a fact observed 
in~\cite{alon_babai_itai}.
If a polynomial $\rndp \in \F_n[x]$ with
degree $c-1$ is chosen uniformly at random,
they observed that
the random variables $\rndp(1),\dots,\rndp(k)$
are $c$-wise independent, and are each uniform in $[n]$.
This allows us to define a distribution on the inputs (the positive test
distribution) that has high agreement with 
$\fnhr$ 
and is easy to
analyze. 
Observe further that, since 
$\abs{\hrfml} \leq n^c$,
the monotone complexity of 
$\fnhr$ 
is at most $2^{O(c \log n)}$.
Later we will choose $c$ to be roughly $n^{1/2}$, and prove that
the monotone complexity of 
$\fnhr$ 
is $2^{\Omega(c)}$.

Finally, 
the restriction 
$\abs{S_P} \geq k/2$ 
is 
a truncation made to
ensure that no minterm of 
$\fnhr$
is very small. 
Otherwise,
if 
$\fnhr$ 
had small minterms,
it might have been a function that almost always outputs~$1$.
Such functions have 
very few maxterms and are therefore computed by a
small CNF. Since we desire 
$\fnhr$ 
to have high complexity, this is an
undesirable property.
The fact that 
$\fnhr$ 
doesn't have small minterms is important
in the proof that 
$\fnhr$ 
almost surely outputs~$0$ in the negative test
distribution (Lemma~\ref{claim:negative_test}).

\begin{remark}[Parameters are now fixed]
    \label{rk:hr_function}
    Formally, the function $\fnhr$ depends on the choice of the parameters
    $c$ and $k$. In other words, for every choice of positive integers $c, k$ such
    that
    $c < k < n$,
    we obtain a different function
    $\fnhr^{(c,k)}$.
    For 
    the rest of Section~\ref{sec:harnik_raz},
    we will let $c$ and $k$ be fixed parameters,
    and we will refer to $\fnhr$ unambiguously,
    always with respect to the fixed parameters $c$ and $k$.
    We will make our choice of $c$ and $k$ explicit in
    Section~\ref{sec:hr_lower_bound},
    but before then we will make no assumptions about $c$ and $k$
    other than $c < k < n$.
\end{remark}

\subsection{Test distributions}
\label{sec:hr_test}

We now define the positive and negative test distributions. 

\begin{definition}[Test distributions]
\label{def:hr_test}
Let
$\rndy \in \blt^n$
be the random variable
which chooses a polynomial $\rndp \in \F_n[x]$
with degree at most $c-1$ uniformly at random,
and maps it into the binary input
$x_{S_{\rndp}} \in \blt^n$.
Let also $\rndn$ be the 
$(1/2)$-biased distribution on $\blt^n$
(i.e., each bit is equal to~$1$ with probablity 
$1/2$, 
independently of all the others).
Equivalently, $\rndn$ is the uniform distribution
on $\blt^n$.
\end{definition}

Harnik and Raz proved that $\fnhr$ outputs~1 on $\rndy$
with high probability.
For completeness, we include their proof.
\begin{lemma}[Claim 4.1 in~\cite{HR00}]
    \label{claim:hr_result}
    We have
    $
    \Pr[\fnhr(\rndy)=1]
    \geq
    1-(k-1)/n.
    $
\end{lemma}
\begin{proof}
    Let $\rndp$ be the polynomial randomly chosen by $\rndy$.
    Call a pair $\set{i,j} \sseq [k]$ with $i \neq j$
    \defx{coinciding} if $\rndp(i) = \rndp(j)$.
    Because the random variables $\rndp(i)$ and $\rndp(j)$ are uniformly
    distributed in $[n]$ and independent for $i \neq j$, we have that
    $\Pr[\rndp(i) = \rndp(j)] = 1/n$ for $i \neq j$.
    Therefore, the expected number 
    $\mathsf{Num}(\rndp)$ 
    of coiciding pairs 
    is 
    $\binom{k}{2}/n$.
    Observe now that 
    $\fnhr(\rndy) = 0$ if and only if
    $\abs{\rndp(1), \dots, \rndp(k)} < k/2$, which occurs only if
    there exists more than $k/2$ coinciding pairs.
    Therefore, by Markov's
    inequality,
    we have
    \begin{equation*}
        \Pr
        \left[ 
            \fnhr(\rndy) = 0
        \right]
        \leq
        \Pr
        \left[ 
            \mathsf{Num}(\rndp)
            > k/2
        \right]
        \leq
        \frac{\binom{k}{2}/n}{k/2}
        =
        \frac{k-1}{n}.
        \qedhere
    \end{equation*}
    \qed
\end{proof}

We now claim that $\fnhr$ also outputs~0 on $\rndn$
with high probability.

\begin{lemma}
    \label{claim:negative_test}
    We have
    $
    \Pr[\fnhr(\rndn)=0]
    \geq
    1-2^{-(k/2-c\cdot\log_2 n)}.
    $
\end{lemma}
\begin{proof}
    Let $x_{\rndA}$ be an input sampled from $\rndn$.
    Observe that $\fnhr(x_{\rndA})=1$ only if there exists
    a minterm $x$ of $\fnhr$ such that $x \leq x_{\rndA}$.
    Since all minterms of $\fnhr$ have Hamming weight at least $k/2$
    and $\fnhr$ has at most $n^c$ minterms,
    we have
    \begin{equation*}
        \Pr[\fnhr(\rndn)=1]
        \leq
        n^c
        \cdot
        2^{-k/2}
        =
        2^{-(k/2-c\cdot\log_2 n)}.
    \end{equation*}
\end{proof}

We will also need the following property about the positive test
distribution.
\begin{lemma}
    \label{lemma:hr_trimmed_prob}
    For every $\ell \leq c$ and $A \sseq [n]$ such that $\abs{A}=\ell$,
    we have
    \begin{equation*}
    \Pr[x_A \leq \rndy]
    \leq
    \left( k/n \right)^{\ell}.
    \end{equation*}
\end{lemma}
\begin{proof}
Recall 
that the distribution $\rndy$ takes a polynomial
$\rndp \in \F_n[x]$
with degree at most $c-1$
uniformly at random 
and returns the binary vector
$x_{\set{\rndp(1), \rndp(2),\dots,\rndp(k)}} \in \blt^n$.
Let $A \in \binom{[n]}{\ell}$ for $\ell \leq c$.
Observe that 
$x_A \leq {\rndy}$
if and only if
$A \sseq \set{\rndp(1), \rndp(2),\dots,\rndp(k)}$.
Therefore, if
$x_A \leq \rndy$,
then there exists
indices $\set{j_1,\dots,j_\ell}$
such that
$\set{\rndp(j_1), \rndp(j_2),\dots,\rndp(j_{\ell})} = A$.
Since $\ell \leq c$, 
we get by the $c$-wise independence of 
$\rndp(1),\dots,\rndp(k)$
that the random variables
$\rndp({j_1}), \rndp({j_2}),\dots,\rndp({j_\ell})$
are independent.
It follows that
\begin{equation*}
    \Pr[
        \set{\rndp(j_1), \rndp(j_2),\dots,\rndp(j_{\ell})} = A
    ]
    =
    \frac{\ell !}{n^\ell}.
\end{equation*}
Therefore, we have
\begin{equation*}
    \Pr[
        x_A \leq \rndy
    ]
    =
    \Pr[ A \sseq \set{\rndp(1), \rndp(2),\dots,\rndp(k)} ]
    \leq
    \binom{k}{\ell}
    \frac{\ell !}{n^\ell}
    \leq
    \left( \frac{k}{n} \right)^{\ell}. \qedhere
\end{equation*}
\end{proof}

\subsection{A closure operator}
\label{sec:hr_closure}

In this section, we describe a closure operator in the lattice of
monotone Boolean functions. We prove that the closure 
of a
monotone Boolean function $f$ is a good approximation for $f$ on 
the negative test distribution~(Lemma~\ref{lemma:hr_error_closure}), and
we give a bound
on the size of the set of minterms of
\emph{closed}
monotone functions.
This bound 
makes use of the 
robust sunflower lemma
(Theorem~\ref{thm:improved_sunflower}),
and is crucial to  bounding errors of approximation
(Lemma~\ref{lemma:hr_error_trimmed}).
Finally, we observe that input functions are closed
(Lemma~\ref{lemma:hr_input_approx}).
From now on, we let
\begin{equation}
    \label{eq:hr_eps}
    \eps := n^{-2c}.
\end{equation}

\begin{definition}[Closed function]
    We say that 
    a monotone function
    $f : \blt^n \to \blt$
    is 
    \emph{closed}
    if, for every
    $A \in \binom{[n]}{\le c}$,
    we have
    \[
        \Pr[\ f(\rndn \vee x_A) = 1\ ] > 
        1 - 
        \eps 
        \ \Longrightarrow\ 
        f(x_A) = 1.
    \]
\end{definition}
This means that for, a closed function,
we always have
$\Pr[f(\rndn \vee x_A) = 1] \notin 
(1-\eps,1)$
when $\abs{A} \leq c$.

\begin{remark}
    [On the parametrization of closedness]
    \label{rk:hr_closed}
    We remark that the definition of a \emph{closed} function
    depends on
    two parameters:
    the parameter $\eps$, defined in~(\ref{eq:hr_eps}),
    and 
    the parameter $c$, used in the construction
    of $\fnhr$ (see Remark~\ref{rk:hr_function}).
    Since 
    both of these parameters
    are
    fixed throughout
    Section~\ref{sec:harnik_raz},
    it is safe to omit 
    them
    without risk of 
    confusion.
    Therefore, we will henceforth say that some function is \emph{closed}
    without any further specification about the parameters.
    However, the reader must bear in mind that, whenever a function is said
    to be \emph{closed},
    the \emph{fixed} parameters $c$ and $\eps$
    are
    in view.
\end{remark}

\begin{definition}[Closure operator]
    \label{def:hr_closure}
    Let $f$ be a monotone Boolean function.
    We denote by $\cl(f)$ the
    unique minimal 
    closed
    monotone Boolean function
    such that
    $f \leq \cl(f)$.
    In other words,
    the function $\cl(f)$
    is the unique 
    closed
    monotone function such that,
    whenever $f \leq g$ and $g$ is monotone and 
    closed,
    we have
    $f \leq \cl(f) \leq g$.
\end{definition}

\begin{remark}
    [On closure]
    \label{rk:hr_closure}
    Note that $\cl(f)$ is well-defined, since the constant Boolean function
    that outputs~$1$ is 
    closed
    and,
    if $f,g$ are both 
    closed
    monotone Boolean
    functions, then so is $f \wedge g$. 
    Furthermore, just as with the definition of closed functions
    (see Remark~\ref{rk:hr_closed}),
    the closure operator $\cl(\cdot)$ depends crucially on
    the 
    parameters $\eps$ and $c$,
    which 
    are 
    fixed throughout Section~\ref{sec:harnik_raz}.
\end{remark}

We now give a bound on the error of approximating $f$ by $\cl(f)$ 
under the distribution $\rndn$.

\begin{lemma}[Approximation by closure]
    \label{lemma:hr_error_closure}
    For every monotone $f: \blt^n \to \blt$,
    we have
    \begin{equation*}
        \Pr
        \left[ 
            f(\rndn) = 0
            \text{ and }
            \cl(f)(\rndn) = 1
        \right]
        \leq
        n^{-c}.
    \end{equation*}
\end{lemma}

\begin{proof}
    We first prove that there exists 
    a positive integer $t$
    and
    sets
    $A_1, \dots, A_t$
    and
    monotone functions
    $h_0, h_1, \dots, h_t : \blt^n \to \blt$ such that
    \begin{enumerate}
        \item 
            $h_0 = f$,
        \item 
            $h_i = h_{i-1} \vee \clqind{A_i}$,
        \item 
            $\Pr[h_{i-1}(\rndn \lor x_{A_i}) = 1] \geq 
            1-\eps$,
        \item 
            $h_t = \cl(f)$.
    \end{enumerate}
    Indeed,
    if $h_{i-1}$ is not closed,
    there exists 
    $A_i \in \binom{[n]}{\leq c}$
    such that
    $\Pr[h_{i-1}(\rndn \lor x_{A_i}) = 1] \geq 
    1-\eps$
    but
    $h_{i-1}(x_{A_i})=0$.
    We let
    $h_i := h_{i-1} \vee \clqind{A_i}$.
    Clearly, we have that $h_t$ is closed,
    and that the value of $t$ is at most the number of 
    subsets of $[n]$
    of size at most $c$. Therefore, we get
    $
        t 
        \leq
        \sum_{j=0}^{c} \binom{n}{j}.
    $
    Moreover, by induction we obtain that $h_i \leq \cl(f)$ for every 
    $i \in [t]$. It follows that $h_t = \cl(f)$.
Now, observe that
    \begin{align*}
        \Pr
        \left[ 
            f(\rndn) = 0
            \text{ and }
            \cl(f)(\rndn) = 1
        \right]
        &\leq
        \sum_{i=1}^t
        \Pr
        \left[ 
            h_{i-1}(\rndn) = 0
            \text{ and }
            h_i(\rndn) = 1
        \right]
        \\&=
        \sum_{i=1}^t
        \Pr
        \left[ 
            h_{i-1}(\rndn) = 0
            \text{ and }
            x_{A_i} 
            \leq
            \rndn 
        \right]
        \\&\leq
        \sum_{i=1}^t
        \Pr
        \left[ 
            h_{i-1}(\rndn \lor x_{A_i}) = 0
        \right]
        \\&\leq
        \eps
        \sum_{j=0}^{c} \binom{n}{j}
        \leq
        n^{-c}.
        \qedhere
    \end{align*}
\end{proof}

We now 
bound
the size of the set of $\ell$-minterms of 
a closed
function. 
This bound 
depends on
the robust sunflower
theorem~(Theorem~\ref{thm:improved_sunflower}).
\begin{lemma}[Closed functions have few minterms]
    \label{lemma:hr_bound_minterms}
    Let $B > 0$ be as in Theorem~\ref{thm:improved_sunflower}.
    If a monotone function $f : \blt^n \to \blt$ is 
    closed,
    then, 
    for all $\ell \in [c]$,
    we have
    \[
        \abs{\mintms_\ell(f)}
        \le 
        (6B c \log n)^\ell
    \]
\end{lemma}

\begin{proof}
    Fix $\ell \in [c]$.
    For convenience, let $p=1/2$ 
    and recall from~(\ref{eq:clq_eps}) that $\eps=n^{-2c}$.
    We will begin by proving that
    $
        \abs{\mintms_\ell(f)}
        \leq
        (B \log(\ell/\eps)/p)^\ell.
    $

    For a contradiction, suppose
    we have
    $
        \abs{\mintms_\ell(f)}
        >
        (B \log(\ell/\eps)/p)^\ell.
    $
    Consider the family
    $\calf := \set{A \in \binom{[n]}{\ell}: x_A \in \mintms_\ell(f)}$.
    Observe that $\abs{\calf} = \abs{\mintms_\ell(f)}$.
    By Theorem~\ref{thm:improved_sunflower},
    there exists a
    $(p,\eps)$-robust sunflower $\calf' \sseq \calf$.
    Let $Y := \bigcap \calf'$ and let $\rndw \sseq_p [n]$.
    We have
    \begin{align*}
        \Pr[f(\rndn \vee x_Y)=1]
        &\geq
        \Pr[\exists x \in \mintms_\ell(f) : x \leq \rndn \vee x_Y]
        \\&=
        \Pr[\exists F \in \calf : F \sseq \rndw \cup Y ]
        \\&\geq
        \Pr[\exists F \in \calf' : F \sseq \rndw \cup Y ]
        \\&>
        1-\eps.
    \end{align*}
    Therefore,
    since $f$ is
    closed,
    we get that
    $f(x_Y)=1$.
    However, since $Y = \bigcap \calf'$,
    there exists $F \in \calf'$ such that $Y \subsetneq F$.
    This is a contradiction, because $x_F$ is a minterm of $f$.
    We conclude that
    \begin{equation*}
        \abs{\mintms_\ell(f)} 
        \leq 
        (B \log(\ell/\eps)/p)^\ell
        \leq
        (2B \log(cn^{2c}))^\ell
        \leq
        (6B c \log n)^\ell.
        \qedhere
    \end{equation*}
\end{proof}

\begin{lemma}[Input functions are 
    closed]
    \label{lemma:hr_input_approx}
    For all $i \in [n]$
    and large enough $n$,
    the Boolean functions
    $\clqind{\set{i}}$
    are 
    closed.
\end{lemma}
\begin{proof}
    Fix $i \in [n]$.
    Let $A \sseq [n]$ be such that $\card{A} \leq c$
    and suppose that 
    $\clqind{\set{i}}(x_A) = 0$.
    Note that 
    $\clqind{\set{i}}(x_A) = 0$
    is equivalent to
    $(x_A)_i = 0$.
    We have
    \begin{equation*}
        \Pr[\clqind{\set{i}}(\rndn \lor x_A)=1]
        =
        \Pr[(\rndn \lor x_A)_i = 1]
        =
        \Pr[\rndn_i = 1]
            = 1/2 \leq 1-n^{-2c} = 1-\eps,
    \end{equation*}
    since $\rndn$ is $(1/2)$-biased (Definition~\ref{def:hr_test})
    and $\eps = n^{-2c}$ (as fixed in~(\ref{eq:clq_eps})).
    Therefore, $\clqind{\set{i}}$ is closed.
\end{proof}

\subsection{Trimmed monotone functions}
\label{sec:hr_trimmed}

In this section, we define a \emph{trimming} operation 
for Boolean functions.
We will bound the probability that a \emph{trimmed} function gives the
correct output on the distribution $\rndy$,
and we will give a bound
on the error of approximating a
Boolean function $f$
by the trimming of $f$ on that same distribution.

\begin{definition}[Trimmed functions]
    We say that a monotone function
    $f \in \blt^n \to \blt$
    is
    \emph{trimmed}
    if 
    all the minterms of $f$ have size at most $c/2$.
    We define the trimming operation $\trim(f)$ as follows:
    \begin{equation*}
        \trim(f)
        :=
        \bigvee_{\ell = 0}^{{c/2}}
        \bigvee_{A \in \mintms_{\ell}(f)}
        \clqind{A}.
    \end{equation*}
\end{definition}
That is, the $\trim$ operation takes out from $f$ all the
minterms of size larger than $c/2$, yielding a
trimmed function.

\begin{remark}
    [Parametrization of $\trim(\cdot)$ and other remarks]
    \label{rk:hr_trim}
    We remark that the definition of trimmed functions
    depends on the choice of the parameter $c$.
    As this parameter is fixed 
    (see Remark~\ref{rk:hr_function}),
    the operator $\trim(\cdot)$ is well-defined.
    Moreover, if all minterms of $f$ have Hamming weight larger
    than $c/2$ (i.e., if $\mintms_{\ell}(f) = \emptyset$ for all 
    $\ell \in \set{0,1,\dots,c/2}$),
    then $\trim(f)$ is the constant function that outputs 0.
    Finally, 
    if $f$ is the constant function $\one$, then
    $\trim(f) = \one$, because $\one$ contains
    a minterm of Hamming weight equal to 0.
\end{remark}

We are now able to bound the
probability that a trimmed Boolean function
gives the correct output on distribution $\rndy$
and give a bound on the 
approximation error of the trimming operation.

\begin{lemma}
[Trimmed functions are inaccurate in the positive distribution]
    \label{lemma:hr_compute_error}
    If a monotone function
    $f \in \blt^n \to \blt$
    is trimmed and $f \neq \one$ (i.e.,\ $f$ is not identically $1$),
    then
    \begin{equation*}
        \Pr
        \left[ f(\rndy) = 1 \right]
        \leq
        \sum_{\ell = 1}^{{c/2}}
        \left( \frac{k}{n} \right)^{\ell}
        \abs{\mintms_\ell(f)}.
    \end{equation*}
\end{lemma}
\begin{proof}
    It suffices to see that,
    since $f$ is trimmed,
    if $f(\rndy) = 1$ and $f \neq \one$
    then
    there exists 
    a minterm $x$ of $f$
    with Hamming weight between $1$ and $c/2$
    such that
    $x \leq \rndy$.
    The result follows from 
    Lemma~\ref{lemma:hr_trimmed_prob}
    and the union bound.
\end{proof}

\begin{lemma}[Approximation by trimming]
    \label{lemma:hr_error_trimmed}
    Let $f \in \blt^n \to \blt$ be a 
    monotone
    function,
    all of whose minterms have Hamming weight at most 
    $c$.
    We have
    \begin{equation*}
        \Pr
        \left[ 
            f(\rndy) = 1
            \text{ and }
            \trim(f)(\rndy) = 0
        \right]
        \leq
        \sum_{\ell = {c/2}}^{c}
        \left( \frac{k}{n} \right)^{\ell}
        \abs{\mintms_\ell(f)}.
    \end{equation*}
\end{lemma}
\begin{proof}
    If we have $f(\rndy) = 1$
    and $\trim(f)(\rndy) = 0$,
    then there was a minterm $x$ of $f$
    with Hamming weight larger than $c/2$ that was removed by the
    trimming process.
    Therefore, since $\abs{x} \leq c$ by assumption,
    the result follows
    from Lemma~\ref{lemma:hr_trimmed_prob}
    and the union bound.
\end{proof}

\subsection{The approximators}
\label{sec:hr_approximators}
Let
$
\cala := \set{\trim(\cl(f)) : 
f : \blt^n \to \blt \text{ is monotone}}.
$
Functions in $\cala$ will be called
\emph{approximators}.
We define
the
\emph{approximating}
operations
$\sqcup, \sqcap: \cala \times \cala \to \cala$
as follows:
for
$f, g \in \cala$,
let
\begin{align*}
    f \sqcup g 
    &:= 
    \trim(\cl(f \vee g)),
    \\
    f \sqcap g 
    &:= 
    \trim(\cl(f \wedge g)).
\end{align*}

We now observe
that every input function 
is an approximator.
Indeed, since every input 
$\clqind{\set{i}}$
is closed and trivially trimmed
(Lemma~\ref{lemma:hr_input_approx}),
we have 
$\trim(\cl(\clqind{\set{i}})) = \trim(\clqind{\set{i}}) =
\clqind{\set{i}}$.
Thus, $\clqind{\set{i}} \in \cala$ for all $i \in [n]$.
Therefore, 
we can replace each gate
of a monotone 
$\set{\vee, \wedge}$-circuit $C$
by its corresponding
approximating gate,
thus obtaining
a
$\set{\sqcup, \sqcap}$-circuit
$C^\cala$
computing
an approximator.

The rationale for choosing this set of approximators is as follows.
By letting approximators be the trimming of a closed function,
we are able to plug the bound on the set of $\ell$-minterms
given by the robust sunflower lemma (Lemma~\ref{lemma:hr_bound_minterms})
on Lemmas~\ref{lemma:hr_compute_error} and~\ref{lemma:hr_error_trimmed},
since the trimming operation can only \emph{reduce} the set of minterms.
Moreover, since trimmings can only help to get a negative answer on the
negative test distribution, we can safely apply
Lemma~\ref{lemma:hr_error_closure} when bounding the errors of
approximation.

\subsection{The lower bound}
\label{sec:hr_lower_bound}

In this section,
we prove that the function
$\fnhr$ requires
monotone circuits of size
$2^{\Omega(c)}$.
By properly choosing $c$ and $k$,
this will imply
the promised $\exp({\Omega(n^{1/2-o(1)})})$
lower bound for the Harnik-Raz function.
First, we fix some parameters.
Choose $B$ as in Lemma~\ref{lemma:hr_bound_minterms}.
Let $T := 18B$.
We also let
\begin{equation*}
    k
    :=
    n^{1/2},
    \quad\quad
    c
    :=
    \frac{1}{T}
    \cdot
    (k/\log n)
    =
    \frac{k}{18B \cdot \log n}.
\end{equation*}
For simplicity, we assume these values are integers.
Note that
$c = \Theta(k / \log n) \ll k$.

\begin{lemma}[Approximators make many errors]
    \label{lemma:hr_approx_err}
    For every approximator $f \in \cala$, we have
    \begin{equation*}
        \Pr[f(\rndy)=1]
        +
        \Pr[f(\rndn)=0]
        \leq
        3/2.
    \end{equation*}
\end{lemma}
\begin{proof}
Let $f \in \cala$.
By definition, there exists 
a closed
function $h$ such that $f = \trim(h)$.
Observe that
$\mintms_\ell(f) \sseq \mintms_\ell(h)$
for every 
$\ell \in [c]$.
From 
Lemma~\ref{lemma:hr_bound_minterms}, we get
    \begin{equation*}
        \abs{\mintms_\ell(h)} 
        \leq
        (6B c \log n)^\ell
        =
        (n/3k)^\ell.
    \end{equation*}
Hence,
applying 
Lemma~\ref{lemma:hr_compute_error},
we obtain that,
if $f \neq \one$,
we have
\begin{equation*}
    \Pr[f(\rndy)=1]
    \leq
    \sum_{\ell=1}^{{c/2}}
    \left( \frac{k}{n} \right)^{\ell}
    \abs{\mintms_\ell(h)}
    \leq
    \sum_{\ell=1}^{{c/2}}
    3^{-\ell}
    \leq
    1/2.
\end{equation*}
Therefore,
for every $f \in \cala$
we have
$
\Pr[f(\rndy) = 1] + \Pr[f(\rndn) = 0] \leq 1 + 1/2 \leq 3/2.
$
\end{proof}

\begin{lemma}[$C$ is well-approximated by $C^\cala$]
    \label{lemma:hr_approx_correct}
    Let $C$ be a monotone circuit.
    We have
    \begin{equation*}
        \Pr[C(\rndy) = 1 \text{ and } C^\cala(\rndy) = 0]
        +
        \Pr[C(\rndn) = 0 \text{ and } C^\cala(\rndn) = 1]
        \leq \size(C) \cdot 2^{-\Omega(c)}.
    \end{equation*}
\end{lemma}
\begin{proof}
    We begin by bounding the approximation errors
    under the distribution $\rndy$.
    We will show that,
    for two approximators $f,g \in \cala$,
    if $f \lor g$ 
    accepts an input from $\rndy$,
    then $f \apor g$
    rejects that input with probability at most $2^{-\Omega(c)}$,
    and that the same holds for
    the approximation
    $f \apand g$.

    First note that, 
    if $f, g \in \cala$,
    then all the minterms of both $f \lor g$ and $f \land g$
    have Hamming weight at most $c$,
    since $f$ and $g$ are trimmed.
    Let now
    $h = \cl(f \lor g)$.
    We have
    $(f \apor g)(x) < (f \lor g)(x)$
    only if
    $\trim(h)(x) < h(x)$.
    Since $h$ is closed,
    we get
    from Lemma~\ref{lemma:hr_bound_minterms} that,
    for all $\ell \in [c]$,
    we have
    \begin{equation*}
        \abs{\mintms_\ell(h)} 
        \leq
        (6B c \log n)^\ell
        =
        (n/3k)^\ell.
    \end{equation*}
    We then
    obtain
    the following inequality
    by 
    Lemma~\ref{lemma:hr_error_trimmed}:
    \begin{align*}
        \Pr
        \left[ 
            (f \lor g)(\rndy) = 1
            \text{ and }
            (f \apor g)(\rndy) = 0
        \right]
        \leq
        \sum_{\ell = {c/2}}^c
        \left( \frac{k}{n} \right)^{\ell}
        \abs{\mintms_\ell(h)}
        \le
        \sum_{\ell = {c/2}}^c
        3^{-\ell}
        =
        2^{-\Omega(c)}.
    \end{align*}
    The same argument shows 
    $
        \Pr
        \left[ 
            (f \wedge g)(\rndy) = 1
            \text{ and }
            (f \apand g)(\rndy) = 0
        \right]
        =
        2^{-\Omega(c)}.
    $
    Since there are $\size(C)$ gates in $C$,
    this implies 
    that
    $
        \Pr[C(\rndy) = 1 \text{ and } C^\cala(\rndy) = 0]
        \leq \size(C) \cdot 2^{-\Omega(c)}.
    $

    To bound the approximation errors under
    $\rndn$,
    note that
    $(f \lor g)(x)=0$
    and
    $(f \apor g)(x) = 1$
    only if
    $
    \cl(f \lor g)(x)
    \neq
    (f \lor g)(x)
    $,
    since trimming a Boolean function 
    cannot decrease
    the probability
    that it rejects an input. 
    Therefore,
    by Lemma~\ref{lemma:hr_error_closure}
    we obtain
    \begin{align*}
        \Pr
        \left[ 
            (f \lor g)(\rndn) = 0
            \text{ and }
            (f \apor g)(\rndn) = 1
        \right]
        \leq
        n^{-c}
        =
        2^{-\Omega(c)}.
    \end{align*}
    The same argument shows 
    $
        \Pr
        \left[ 
            (f \land g)(\rndn) = 0
            \text{ and }
            (f \apand g)(\rndn) = 1
        \right]
        =
        2^{-\Omega(c)}.
    $
    Once again, doing this approximation for every gate in $C$ 
    allows us to conclude
    $
        \Pr[C(\rndn) = 0 \text{ and } C^\cala(\rndn) = 1]
        \leq \size(C) \cdot 2^{-\Omega(c)}.
    $
    This finishes the proof.
\end{proof}

\begin{theorem}
    \label{thm:hr_lower_bound}
    Any monotone circuit computing $\fnhr$ has size
    $2^{\Omega(c)} = 2^{\Omega (n^{1/2}/\log n)}$.
\end{theorem}
\begin{proof}
    Let $C$ be a monotone circuit computing $\fnhr$.
    Since $k/2 - c \log_2 n = \Omega(k)$ and
    $k \ll n$,
    for large enough $n$
    we obtain from
    Lemmas~\ref{claim:hr_result} and \ref{claim:negative_test}
    that
    \begin{equation*}
        \Pr[\fnhr(\rndy)=1]
        + 
        \Pr[\fnhr(\rndn)=0]
        \geq 
        2-(k-1)/n-2^{-(k/2-c\log_2 n)}
        \geq
        9/5.
    \end{equation*}
    We then obtain from Lemmas~\ref{lemma:hr_approx_err}
    and~\ref{lemma:hr_approx_correct}:
    \begin{align*}
        9/5
        &\leq
        \Pr[\fnhr(\rndy)=1]
        + 
        \Pr[\fnhr(\rndn)=0]
        \\&
        \leq
        \Pr[C(\rndy) = 1 \text{ and } C^\cala(\rndy) = 0]
        +
        \Pr[C^\cala(\rndy)=1]
        \\&
        +
        \Pr[C(\rndn) = 0 \text{ and } C^\cala(\rndn) = 1]
        +
        \Pr[C^\cala(\rndn)=0]
        \\&
        \leq
        3/2
        +
        \size(C)2^{-\Omega(c)}.
    \end{align*}
    This implies
    $\size(C) = 2^{\Omega(c)}$.
\end{proof}

\subsection{Are better lower bounds possible with robust sunflowers?}
\label{sec:hr_discussion}

In this section, we allow some degree of imprecision
for the sake of brevity and clarity, 
in order to highlight
the main technical ideas of the proof.

A rough outline of how we just proved 
Theorem~\ref{thm:hr_lower_bound} 
is as follows.
First, we noted that the minterms of $\fnhr$ are ``well-spread''.
This is Lemma~\ref{lemma:hr_trimmed_prob},
which states that
the probability that a fixed set $A \sseq [n]$
is contained in a random minterm\footnote{Here, ``random minterm'' means an
input from the distribution $\rndy$, which correlates highly with the
minterms of $\fnhr$.} of $\fnhr$
is at most 
$r^{\card{A}}$, 
where $r = k/n$.
Moreover, we observed that $\fnhr$ outputs $0$ with high probability
in a $p$-biased distribution (Lemma~\ref{claim:negative_test}),
where $p=1/2$.

In the rest of the proof, we roughly showed how this implies that 
DNFs of size 
approximately 
$s = c^{c/2}$ 
and width $w = c/2$
\emph{cannot} approximate $\fnhr$
(Lemma~\ref{lemma:hr_approx_err}).\footnote{
    Formally, our approximators
    have at most $O(c \log n)^\ell$ terms of width $\ell$
    (Lemma~\ref{lemma:hr_bound_minterms}),
    and no terms of width larger than $c/2$ (by trimming).
}
We also observed that 
we can
approximate the $\lor$ and $\land$
of width-$w$, size-$s$ DNFs
by another width-$w$, size-$s$ DNF,
bounding the error of approximation by
$r^{c/2} \cdot c^{c/2}$.
This was proved by noting that
conjunctions of width $c/2$
accept a positive input with probability at most $r^{c/2}$,
and there are at most $c^{c/2}$ of them.
When $c \approx k \approx \sqrt{n}$,
we have $(rc)^{c/2} = 2^{-\Omega(c)}$,
and thus 
\emph{we can} 
approximate circuits of size
$2^{o(c)}$ with width-$w$, size-$s$ DNFs
(Lemma~\ref{lemma:hr_approx_correct}).
This yields the lower bound.

There are two essential numerical components in the proof.
First, the ``spreadness rate'' of the function $\fnhr$.
A simple counting argument can show that the upper bound of
$(k/n)^{\card{A}}$ to the probability
$\Pr[x_A \leq \rndy]$ is nearly best possible
when 
the support of $\rndy$ is contained in
$\blt^n_{= k}$ and $k = o(n)$.
So this can hardly be improved with the choice of another Boolean function.
Secondly, 
the bounds for the size and width of the DNF approximators
come from the robust sunflower lemma
(Theorem~\ref{thm:improved_sunflower}),
which was used to employ the approximation method
on $p$-biased distributions.
Since the bound of Theorem~\ref{thm:improved_sunflower} is essentially
best possible as well,
as observed in~\cite{alweiss2019improved},
we cannot hope to get better approximation bounds on a $p$-biased
distribution from sunflowers.
Therefore, there does not seem to be much room for getting better
lower bounds for monotone circuits
using the classical approximation method with 
sunflowers,
if we use $p$-biased distributions.
To get beyond 
$2^{\Omega(\sqrt{n})}$,
another approach
seems to be required.

\section{Lower Bound for $\kclq$}
\label{sec:clique}

Recall that
the Boolean function
$\kclq : \{0,1\}^{\binom{n}{2}} \to
\{0,1\}$ receives a graph on $n$ vertices as an input
and outputs a $1$ if this graph contains a clique on $k$ vertices. 
In this section, we prove an $n^{\Omega(\delta^2 k)}$ lower bound on the
monotone circuit size of $\kclq$ for $k \leq n^{(1/3)-\delta}$.

We note that the 
first superpolynomial lower bound for the monotone circuit complexity
of $\kclq$
was given by
Razborov
\cite{razborov_boolean_85}, who proved
a $n^{\Omega(k)}$ lower bound 
for $k \leq \log n$.
Soon after, 
Alon and Boppana 
\cite{alon_boppana}
proved
a $n^{\Omega(\sqrt k)}$
for $\kclq$
when $k \le n^{2/3 - o(1)}$.
This exponential lower bound was
better than Razborov's, as it could be applied to
a larger range of $k$,
but it was short of the obvious upper bound of
$n^{O(k)}$.
Our result finally closes that gap,
by proving that the monotone complexity of $\kclq$
is $n^{\Theta(k)}$ even for large $k$.

As in Section~\ref{sec:harnik_raz},
we will follow the approximation method.
However, instead of using sunflowers as in
\cite{razborov_boolean_85,alon_boppana} or robust sunflowers as in
\cite{rossman_kclq},
we introduce a notion of \emph{clique-sunflowers}
and employ it to bound the errors of approximation.

\subsection{Notation for this section}

In this section, we will often refer to graphs on $n$ vertices
and Boolean strings in $\blt^{\binom{n}{2}}$ interchangeably.
For $A \subseteq [n]$, 
let $K_A$ be the graph on $n$ vertices
with a clique
on $A$ and no other edges.
When $\card{A} \leq 1$, the graph $K_A$ is the empty graph with $n$ vertices
and 0 edges (corresponding to the Boolean string
    all of which $\binom{n}{2}$ entries are equal to 0.)
The \emph{size} of $K_A$ is $\card{A}$.
Let also
$\clqind{A} : \blt^{\binom{n}{2}} \to \blt$
denote the indicator function of containing $K_A$,
which satisfies
\begin{equation*}
    \clqind{A}(G) = 1
    \iff
    K_A \sseq G.
\end{equation*}
Functions of the forms 
$\clqind{A}$
are called
\defx{clique-indicators.}
Moreover, if $\abs{A} = \ell$,
we say that
$\clqind{A}$
is a
clique-indicator of \emph{size} equal to $\ell$.
When $\card{A} \leq 1$, the function $\clqind{A}$
is the constant function $\one$.

For $p \in (0,1)$,
we denote by $\mb G_{n,p}$ the
Erd\H{o}s-R\'enyi random graph,
a random graph on $n$ vertices
in which each edge appears independently with probability $p$.

Let $f : \blt^{\binom{n}{2}} \to \blt$ be monotone
and suppose $\ell \in \set{1,\dots, \delta k}$.
We define
\[
    \mintms_\ell(f) \defeq 
    \{A \in \textstyle\binom{[n]}{\ell} : 
    f(K_A) = 1 
    \text{ and } 
    f(K_{A\setminus\{a\}}) = 0\text{ for all } a \in A\}.
\]
Elements of $\mintms_\ell(f)$ are called 
\emph{$\ell$-clique-minterms} of $f$.

\subsection{Clique-sunflowers}

Here we introduce the notion of
\emph{clique-sunflowers},
which is analogous to that of robust sunflowers
for ``clique-shaped'' set systems.

\begin{definition}[Clique-sunflowers]
    \label{def:clique_sunflower}
    Let $\eps, p \in (0,1)$.
    Let $\cals$ be a family of subsets of $[n]$
    and let 
    $Y := \bigcap \cals$.
    The family $\cals$ is called a
    $(p, \eps)$-\emph{clique-sunflower}
    if
    \begin{equation*}
        \Pr
        \left[ 
            \exists A \in S:
            K_A 
            \sseq 
            \mb G_{n,p} \cup K_Y
        \right]
        >
        1-\eps.
    \end{equation*}
    Equivalently,
    the family
    $\cals$
    is a clique-sunflower
    if
    the family
    $\set{K_A : A \in \cals} \sseq \binom{[n]}{2}$
    is a $(p,\eps)$-robust sunflower,
    since $K_A \cap K_B = K_{A \cap B}$.
\end{definition}

Though clique-sunflowers may seem similar to regular sunflowers,
the importance of this definition is that it allows us to explore the
``clique-shaped'' structure of the sets of the family, and thus
obtain an asymptotically better upper bound on the size of sets that
do not contain a clique-sunflower.

\begin{lemma}[Clique-sunflower lemma]
    \label{lemma:clique_sunflower}
    Let $\eps < e^{-1/2}$
    and
    let 
    $\cals \sseq \binom{[n]}{\ell}$.
    If
    the family $\cals$
    satisfies
    $\abs{\cals} >
    \ell!(2\ln(1/\eps))^\ell(1/p)^{\binom{\ell}{2}}$,
    then 
    $\cals$ contains
    a
    $(p,\eps)$-clique-sunflower.
\end{lemma}

Observe that, whereas the bounds for ``standard'' robust sunflowers
(Theorems~\ref{thm:approx_sunflower} and \ref{thm:improved_sunflower})
would give us an exponent of $\binom{\ell}{2}$
on the $\log(1/\eps)$ factor,
Lemma \ref{lemma:clique_sunflower} give us only an $\ell$ at the exponent. 
As we shall see,
this is asymptotically better for our choice of parameters.

We defer the proof of Lemma~\ref{lemma:clique_sunflower} 
to Section~\ref{sec:proof_clqsf}.
The proof is based on an application of Janson's inequality
\cite{janson1990poisson}, as in the original robust sunflower
lemma of \cite{rossman_kclq} (Theorem~\ref{thm:approx_sunflower}).

\subsection{Test distributions}

We now define the positive and negative test distributions.
First, we fix some parameters
that will be used throughout the proof.
Fix $\delta \in (0,1/3)$.
Let
\begin{equation}
    \label{eq:clq_p}
    k = n^{1/3-\delta}
    \quad\text{ and }\quad
    p := n^{-2/(k-1)}.
\end{equation}
For simplicity, we will assume from now on that
$\delta k$ and $\delta k/2$ are integers.

\begin{remark}[Parameters are now fixed]
    \label{rk:clq_parameters}
    From now on until the end of Section~\ref{sec:clq_lb},
    the symbols $p, \delta$ and $k$ refer to fixed parameters,
    and will always unambiguously refer to the values 
    just fixed. 
    This will only change in Section~\ref{sec:proof_clqsf},
    which is independent of the proof of the lower bound for $\kclq$,
    and 
    in which we
    will permit ourselves to reuse some of these
    symbols for other purposes.
    This means that, whenever $p, \delta$ and $k$ appear in the following
    discussion, the reader must bear in mind that
    $p=n^{-2/(k-1)}$, $\delta$ is a fixed number inside $(0,1/3)$
    and $k$ is 
    fixed to be $k=n^{1/3-\delta}$.
\end{remark}

We observe that the probability that $\mb G_{n,p}$ has a $k$-clique is
bounded away from~$1$.
\begin{lemma}
    \label{lemma:clique_random_graph}
    We have
    $\Pr[\ \mb G_{n,p} \text{ contains a $k$-clique}\ ] \le 3/4$.
\end{lemma}

\begin{proof}
    There are $\binom{n}{k} \le (en/k)^k$ potential
    $k$-cliques, each present in $\mb G_{n,p}$ with probability
    $p^{\binom{k}{2}} = %
    n^{-k}$.
    By a union bound,
    we have
    $\Pr[\ \mb G_{n,p} \text{ contains a
    $k$-clique}\ ] 
    \le (e/k)^k \le (e/3)^3 
    \le 3/4$.
\end{proof}

\begin{definition}
    \label{def:clq_test}
Let $\rndy$ be the uniform random graph chosen from all possible $K_A$,
where $\card{A} = k$.
In other words, the distribution $\rndy$ samples a random minterm of
$\kclq$.
We call $\rndy$ the
\emph{positive test distribution}.
Let also
$\rndn := \mb G_{n,p}$.
We call $\rndn$ the
\emph{negative test distribution}.
\end{definition}

From Lemma~\ref{lemma:clique_random_graph}, we easily obtain the
following corollary.

\begin{corollary}
    \label{cor:clique_random_graph}
    We have
    $
        \Pr[\kclq(\rndy)=1]
        + 
        \Pr[\kclq(\rndn)=0]
        \geq
        5/4.
    $
\end{corollary}

We now prove an analogous result to that of
Lemma~\ref{lemma:hr_trimmed_prob},
which shows that the positive distribution $\rndy$
is unlikely to contain a large fixed clique.

\begin{lemma}
    \label{lemma:clq_spread}
    For every $\ell \leq k$ and $A \sseq [n]$ such that $\abs{A}=\ell$,
    we have
    \begin{equation*}
    \Pr[K_A \leq \rndy]
    \leq
    \left( k/n \right)^{\ell}.
    \end{equation*}
\end{lemma}
\begin{proof}
    The distribution $\rndy$
    samples a set $\rndB$ uniformly at random from
    $\binom{[n]}{k}$
    and returns the graph $K_{\rndB}$.
    Note that $K_A \sseq K_{\rndB}$
    if and only if
    $A \sseq \rndB$.
    We have
    \begin{equation*}
    \Pr[K_A \leq \rndy]
    =
    \Pr[A \sseq \rndB]
    =
    \frac{\binom{n-k}{k-\ell}}{\binom{n}{k}}
    \leq
    \left( \frac{k}{n} \right)^{\ell}.
    \qedhere
    \end{equation*}
\end{proof}

\subsection{A closure operator}
\label{sec:clique_closed}

As in Section~\ref{sec:hr_closure},
we define here a closure operator in the lattice of monotone
Boolean functions. We will again prove that the closure of a function
will be a good approximation for it on the negative test distribution.
However, unlike Section~\ref{sec:hr_closure},
instead of bounding the set of minterms, we will bound the set of
``clique-shaped'' minterms, as we shall see.
Finally, we will observe that input functions are also closed.
Henceforth, we fix the error parameter
\begin{equation}
    \label{eq:clq_eps}
    \eps := n^{-k}.
\end{equation}

\begin{definition}[Closed functions]
    We say that $f \in \blt^{\binom{n}{2}} \to \blt$
    is 
    \emph{closed}
    if, for every 
    $A \sseq [n]$
    such that $\abs{A} \in \set{2, \dots, \delta k}$,
    we have
    \[
        \Pr[\ f(\rndn \lor K_A) = 1\ ] > 1 - \eps 
        \ \Longrightarrow\ 
        f(K_A) = 1.
    \]
\end{definition}

\begin{remark}
    [On the parametrization of closedness]
    \label{rk:clq_closed}
    Similarly to the Harnik-Raz case
    (see Remark~\ref{rk:hr_closed}),
    the definition of a \emph{closed} function
    depends on
    three parameters:
    the probability $p$, which controls the distribution $\rndn$
    (as discussed
    in Definition~\ref{def:clq_test}),
    the parameter $\eps$, defined in~(\ref{eq:clq_eps}),
    and 
    the parameter $k$.
    Since 
    all of these three 
    parameters 
    are 
    fixed until the end of
    Section~\ref{sec:clq_lb}~(see Remark~\ref{rk:clq_parameters}),
    and no other reference to closed functions will be made after that,
    it is safe to omit 
    them
    without risk of 
    confusion.
    Therefore, we will henceforth say that some function is \emph{closed}
    without any further specification about the parameters.
    However, the reader must bear in mind that, whenever a function is said
    to be \emph{closed},
    the \emph{fixed} parameters $p, \eps$ and $k$ are
    in view.
\end{remark}

\begin{remark}
    [Definitions of closedness compared]
    Definition~\ref{def:clq_closure} bears great resemblance
    to Definition~\ref{def:hr_closure},
    which also talks about a notion of \emph{closed monotone functions}
    in the context of lower bounds for the function of Harnik and Raz.
    Apart from the different parametrizations,
    the main difference between those two definitions 
    is that,
    whereas Definition~\ref{def:hr_closure}
    looks into \emph{all} inputs of Hamming weight at most $c$,
    here we only care about \emph{clique-shaped} inputs
    of size at most $\delta k$.
\end{remark}

As before, we can define the closure of a monotone Boolean function
$f$.

\begin{definition}[Closure operator]
    \label{def:clq_closure}
    Let $f$ be a monotone Boolean function.
    We denote by $\cl(f)$ the
    unique minimal 
    closed
    monotone Boolean function
    such that
    $f \leq \cl(f)$.
\end{definition}

\begin{remark}[On closure]
    \label{rk:clq_closure}
We note again that $\cl(f)$ is well-defined 
(the same arguments of Remark~\ref{rk:hr_closure} apply here)
and remark that its definition also depends on the parameters
$p, \eps$ and $k$ (see Remark~\ref{rk:clq_closed}),
which are fixed throughout the proof,
and therefore can be safely omitted.
\end{remark}

\begin{lemma}[Approximation by closure]
    \label{lemma:clique_error_closure}
    For every monotone $f : \blt^{\binom{n}{2}} \to \blt$,
    we have
    \begin{equation*}
        \Pr
        \left[ 
            f(\rndn) = 0
            \text{ and }
            \cl(f)(\rndn) = 1
        \right]
        \le
        n^{-(2/3)k}.
    \end{equation*}
\end{lemma}
\begin{proof}
    We repeat the same argument as that of
    Lemma~\ref{lemma:hr_error_closure}.
    Since there are at most $n^{\delta k}$
    graphs $K_A$ such that $\card{A} \leq \delta k$
    and $\eps = n^{-k}$,
    the final bound then becomes
    $n^{-k} \cdot n^{\delta k} \leq n^{-(2/3)k}$.
\end{proof}

By employing the clique-sunflower lemma
(Lemma~\ref{lemma:clique_sunflower}),
we are able to bound the set of $\ell$-clique-minterms of closed
monotone functions.

\begin{lemma}[Closed functions have few minterms]
    \label{lemma:bound_minterms}
    If a monotone function
    $f : \blt^{\binom{n}{2}} \to \blt$ is 
    closed,
    then, for all $\ell \in \set{2,\dots, \delta k}$, we have
    \[
        \abs{\mintms_\ell(f)}
        &\le 
        n^{2\ell/3}.
    \]
\end{lemma}
\begin{proof}
    Recall that $p = n^{-2/(k-1)}$ 
    and
    $\eps = n^{-k}$
    (see~(\ref{eq:clq_p}) and (\ref{eq:clq_eps})).
    Applying the same strategy of 
    Lemma~\ref{lemma:hr_bound_minterms},
    replacing the application of
    Theorem~\ref{thm:improved_sunflower} (robust sunflower theorem)
    by Lemma~\ref{lemma:clique_sunflower} (clique-sunflower lemma),
    we obtain
    \[
        \abs{\mintms_\ell(f)}
        &\leq
        \ell!(2\log(1/\eps))^\ell(1/p)^{\binom{\ell}{2}}
        \leq
        (2\ell k \log n)^\ell 
        \cdot 
        {p^{-\binom{\ell}{2}}}
        \\&
        \leq
        (2\delta k^2 \log n)^\ell
        \cdot
        {n^{2\binom{\ell}{2}/(k-1)}}
        \leq
        (n^{2/3-2\delta} \log n)^\ell
        \cdot
        n^{\delta \ell}
        \leq
        n^{2\ell/3}.
        \qedhere
    \]
\end{proof}

\begin{lemma}[Input functions are 
    closed]
    \label{lemma:clq_input_approx}
    Let $i,j \in [n]$ be such that
    $i \neq j$.
    For large enough $n$,
    the Boolean function
    $\clqind{\set{i,j}}$
    is
    closed.
\end{lemma}
\begin{proof}
    Fix $i,j \in [n]$ such that $i \neq j$.
    Let $A \sseq [n]$ be such that $\card{A} \leq \delta k$
    and suppose that 
    $\clqind{\set{i,j}}(K_A) = 0$.
    Note that $\clqind{\set{i,j}}(K_A) = 0$
    is equivalent to
    $\set{i,j} \not\sseq A$.
    This implies that
    $\set{i,j}$ is an edge of $\mb \rndn \cup K_A$
    if and only if
    $\set{i,j}$ is an edge of $\rndn$.
    Therefore, 
    we have
    \begin{align*}
        \Pr[\clqind{\set{i,j}}(\rndn \lor K_A)=1]
        &=
        \Pr[\clqind{\set{i,j}}(\rndn)=1]
        \\&
        =
        \Pr[\set{i,j} \text{ is an edge of } \mb G_{n,p}]
        \\&
        =
        n^{-2/(k-1)},
    \end{align*}
    since $\rndn = \mb G_{n,p}$ and $p = n^{-2/(k-1)}$
    (see 
    (\ref{eq:clq_p}),
    Remark~\ref{rk:clq_parameters}
    and Definiton~\ref{def:clq_test}).
    It now suffices to show that, for large enough $n$, we have
    $p \leq 1-\eps=1-n^{-k}$ (recall from~(\ref{eq:clq_eps}) that
    $\eps = n^{-k}$).

    For convenience, let $\alpha = 1/3-\delta$. Note that $k=n^{\alpha}$.
    For large enough $n$, we have
    \begin{equation*}
        \frac{2 \cdot \log n}{n^{\alpha}-1} 
        \geq
            n^{-n^\alpha}
            +
            n^{-2n^\alpha}.
    \end{equation*}
    Using the inequality
    $\log(1-x) \geq -x -x^2$ for $x \in [0,1/2]$,
    we get 
    \begin{equation*}
        \frac{2 \cdot \log n}{k-1} 
        =
        \frac{2 \cdot \log n}{n^{\alpha}-1} 
        \geq
            n^{-n^\alpha}
            +
            n^{-2n^\alpha}
        \geq
        -\log(1-n^{-n^\alpha})
        =
        -\log(1-n^{-k}).
    \end{equation*}
    Therefore, we have
    \begin{equation*}
        n^{-2/(k-1)}
        \leq
        1-n^{-k},
    \end{equation*}
    and we conclude that
    $\clqind{\set{i,j}}$ is closed.
\end{proof}

\subsection{Trimmed monotone functions}
In this section, we define again a trimming operation for Boolean
functions and prove analogous bounds to that of
Section~\ref{sec:hr_trimmed}.

\begin{definition}[Clique-shaped and trimmed functions]
    We say that a function
    $f : \blt^{\binom{n}{2}} \to \blt$
    is
    \emph{clique-shaped}
    if, for every minterm $x$ of $f$,
    there exists $A \sseq [n]$
    such that $x = K_A$.
    Moreover, we say that $f$
    is
    \emph{trimmed}
    if 
    $f$ is clique-shaped and
    all the clique-minterms of $f$ have size at most~$\delta k/2$.
    For a clique-shaped function $f$, we define 
    the trimming operation $\trim(f)$ as follows:
    \begin{equation*}
        \trim(f)
        :=
        \bigvee_{\ell = 1}^{\delta k/2}%
        \bigvee_{A \in \mintms_{\ell}(f)}
        \clqind{A}.
    \end{equation*}
    That is, the $\trim$ operation takes out from $f$ all the
    clique-indicators of size larger than %
    $\delta k/2$, yielding a
    trimmed function.
\end{definition}

\begin{remark}[Parametrization of $\trim(\cdot)$ and other remarks]
    \label{rk:clq_trim}
    Analogously
    to the Harnik-Raz case (see Remark~\ref{rk:hr_trim}),
    the definition of trimmed functions
    depends on the choice of the parameters $\delta$ and $k$.
    As these parameters are fixed
    (see Remark~\ref{rk:clq_parameters}),
    the operator $\trim(\cdot)$ is well-defined.
    Moreover, if all clique-minterms of $f$ have size larger
    than $\delta k/2$ 
    (i.e., if $\mintms_{\ell}(f) = \emptyset$ for all $\ell \in
    [\delta k/2]$),
    then $\trim(f)$ is the constant function that outputs 0.
    Finally, 
    if $f$ is the constant function $\one$, then
    $\trim(f) = \one$, because $\one$ contains
    a clique-minterm of size equal to 1 
    (a clique containing one vertex and no edges).
\end{remark}

Imitating the proofs of
Lemmas~\ref{lemma:hr_compute_error} and~\ref{lemma:hr_error_trimmed},
replacing Lemma~\ref{lemma:hr_trimmed_prob}
by
Lemma~\ref{lemma:clq_spread},
we may now obtain the following lemmas.

\begin{lemma}
[Trimmed functions are inaccurate in the positive distribution]
    \label{lemma:clique_shaped_error}
    If a monotone function
    $f : \blt^{\binom{n}{2}} \to \blt$ 
    is a trimmed
    clique-shaped function 
    such that $f \neq \one$,
    then
    \begin{equation*}
        \Pr
        \left[ f(\rndy) = 1 \right]
        \leq
        \sum_{\ell = 2}^{\delta k/2}%
        \left( \frac{k}{n} \right)^{\ell}
        \abs{\mintms_\ell(f)}.
    \end{equation*}
\end{lemma}

\begin{lemma}[Approximation by trimming]
    \label{lemma:clique_error_trimmed}
    Let 
    $f : \blt^{\binom{n}{2}} \to \blt$ 
    be a clique-shaped 
    monotone
    function,
    all of whose clique-minterms have size at most 
    $\delta k$.
    We have
    \begin{equation*}
        \Pr
        \left[ 
            f(\rndy) = 1
            \text{ and }
            \trim(f)(\rndy) = 0
        \right]
        \leq
        \sum_{\ell={\delta k/2}}^{\delta k}%
        \left( \frac{k}{n} \right)^{\ell}
        \abs{\mintms_\ell(f)}.
    \end{equation*}
\end{lemma}

\subsection{Approximators}

Similarly as in Section~\ref{sec:hr_approximators},
we will consider a set of
\emph{approximators} $\cala$.
Let
$$\cala := \{\trim(\cl(f)) : 
    f \in \blt^{\binom{n}{2}} \to \blt 
    \text{ is monotone and
clique-shaped}\}.$$
Functions in $\cala$ are called
\emph{approximators}.
Note that every function in $\cala$ is clique-shaped
and is the trimming of a closed function.
Moreover, observe that every edge-indicator $\clqind{\set{u,v}}$ belongs to
$\cala$, since 
every edge-indicator is
closed by Lemma~\ref{lemma:clq_input_approx}. 

Let
$f, g \in \cala$
such that
$f = \bigvee_{i=1}^t \clqind{A_i}$
and
$g = \bigvee_{j=1}^s \clqind{B_j}$.
We define
$\bigwedge(f,g) := \bigvee_{i=1}^t \bigvee_{j=1}^s \clqind{A_i \cup B_j}$.
We also define
operations
$\sqcup, \sqcap: \cala \times \cala \to \cala$
as follows:
\begin{align*}
    f \sqcup g 
    &:= 
    \trim(\cl(f \vee g)),
    \\
    f \sqcap g 
    &:= 
    \trim\left(
        \cl
        \left(
            \bigwedge(f,g)
        \right)
    \right).
\end{align*}

    It's easy to see that, if $f,g \in \cala$,
    then $f \sqcup g \in \cala$.
    To see that $f \sqcap g \in \cala$,
    note that
    $\bigwedge(f,g)$ is also a monotone clique-shaped function.

\begin{remark}[Reason for definition of $\sqcap$]
    \label{rk:clq_approx_and}
    The reason for defining $\sqcap$ in that way is as follows.
    First observe that $f \land g = \bigvee_{i = 1}^t \bigvee_{j=1}^s (\clqind{A_i}
    \land \clqind{B_j})$.
    We simply replace each $\clqind{A_i} \cap \clqind{B_j}$ with
    $\clqind{A_i \cup B_j}$,
    thus obtaining $f \sqcap g$.
    In general, since $\clqind{A_i \cup B_j}$ is a larger conjunction
    than $\clqind{A_i} \land \clqind{B_j}$, we have
    $\bigwedge(f,g) \leq f \land g$.
    However, note that, for every $A \sseq [n]$, we have
    $\bigwedge(f,g)(K_A) = (f \land g)(K_A)$.
    Thus, the transformation from
    $f \land g$ to
    $\bigwedge(f,g)$ incurs no 
    mistakes in the positive distribution $\rndy$.
\end{remark}

If $C$ is a monotone
$\set{\vee, \wedge}$-circuit,
let
$C^\cala$
be the corresponding
$\set{\sqcup, \sqcap}$-circuit,
obtained by replacing each
$\lor$-gate by a $\sqcup$-gate,
and 
each
$\land$-gate by an $\sqcap$-gate.
Note that $C^\cala$ computes
an approximator.

\subsection{The lower bound}
\label{sec:clq_lb}

In this section we obtain
the lower bound for the clique function.
Recall that $k = n^{1/3-\delta}$.
We will prove that
the monotone complexity of 
$\kclq$ is
$n^{\Omega(\delta^2 k)}$.

Repeating the same arguments of
Lemmas~\ref{lemma:hr_approx_err}
and~\ref{lemma:hr_approx_correct},
we obtain the following analogous lemmas.
\begin{lemma}[Approximators make many errors]
    \label{lemma:clique_approx_err}
    For every $f \in \cala$,
    we have
    \begin{equation*}
        \Pr[f(\rndy)=1]
        +
        \Pr[f(\rndn)=0]
        \leq
        1+o(1).
    \end{equation*}
\end{lemma}
\begin{proof}
Let $f \in \cala$.
By definition, there exists 
a closed
function $h$ such that $f = \trim(h)$.
Observe that
$\mintms_\ell(f) \sseq \mintms_\ell(h)$
for every 
$\ell \in \set{2,\dots,\delta k/2}$.
By
Lemmas~\ref{lemma:bound_minterms}
and~\ref{lemma:clique_shaped_error},
if $f \in \cala$ is such that $f \neq \one$,
then
\begin{equation*}
    \Pr[f(\rndy)=1]
    \leq
    \sum_{\ell=2}^{\delta k/2}
    \left( \frac{k}{n} \right)^{\ell}
    \abs{\mintms_\ell(h)}
    \leq
    \sum_{\ell=2}^{\delta k/2}
    \left(\frac{k}{n^{1/3}}\right)^\ell
    \leq
    \sum_{\ell=2}^{\delta k/2}
    n^{-\delta \ell}
    =
    o(1).
\end{equation*}
Therefore,
for every $f \in \cala$
we have
$
\Pr[f(\rndy) = 1] + \Pr[f(\rndn) = 0] \leq 1 + o(1).%
$
\end{proof}

\begin{lemma}[$C$ is well-approximated by $C^\cala$]
    \label{lemma:clique_approx_correct}
    Let $C$ be a monotone circuit.
    We have
    \begin{equation*}
        \Pr[C(\rndy) = 1 \text{ and } C^\cala(\rndy) = 0]
        +
        \Pr[C(\rndn) = 0 \text{ and } C^\cala(\rndn) = 1]
        \leq \size(C)\cdot O(n^{-\delta^2 k / 2}).
    \end{equation*}
\end{lemma}

\begin{proof}
    To bound the approximation errors
    under the distribution $\rndy$,
    first note that, 
    if $f, g \in \cala$,
    then all the clique-minterms of both $f \lor g$ and $f \land g$
    have size at most $\delta k$.
    Moreover,
    if 
    $(f \lor g)(x)=1$
    but 
    $(f \apor g)(x) = 0$,
    then
    $
    \trim(\cl(f \lor g)(x))
    \neq
    \cl(f \lor g)(x)
    $.
    Therefore,
    we obtain by Lemmas~\ref{lemma:bound_minterms}
    and~\ref{lemma:clique_error_trimmed}
    that, for $f, g \in \cala$,
    we have
    \[
        \Pr
        \left[ 
            (f \lor g)(\rndy) = 1
            \text{ and }
            (f \apor g)(\rndy) = 0
        \right]
        &\leq
        \sum_{\ell = \delta k /2}^{\delta k}
        \left( \frac{k}{n} \right)^{\ell}
        \abs{\mintms_\ell(\cl(f \lor g))}
        \\&\le
        \sum_{\ell = \delta k /2}^{\delta k}
        n^{-\delta \ell}
        =
        O(n^{-\delta^2 k / 2}).
    \]
    As observed in Remark~\ref{rk:clq_approx_and},
    we have $\bigwedge(f,g)(\rndy) = (f \land g)(\rndy)$.
    Thus, once again, the only approximation mistakes incurred
    by changing a $\land$-gate for a $\sqcap$-gate
    comes from the trimming operation. 
    Again, we conclude
    $$
        \Pr
        \left[ 
            (f \wedge g)(\rndy) = 1
            \text{ and }
            (f \apand g)(\rndy) = 0
        \right]
        =
        O(n^{-\delta^2 k / 2}),
    $$
    which implies 
    \begin{equation*}
        \Pr[C(\rndy) = 1 \text{ and } C^\cala(\rndy) = 0]
        \leq \size(C) \cdot O(n^{-\delta^2 k / 2}).
    \end{equation*}

    Similarly, to bound the approximation errors under
    $\rndn$,
    note that
    $(f \lor g)(x)=0$
    and
    $(f \apor g)(x) = 1$
    only if
    $
    \cl(f \lor g)(x)
    \neq
    (f \lor g)(x)
    $.
    Therefore,
    we obtain by Lemma~\ref{lemma:clique_error_closure}
    that, for $f, g \in \cala$,
    we have
    \begin{align*}
        \Pr
        \left[ 
            (f \lor g)(\rndn) = 0
            \text{ and }
            (f \apor g)(\rndn) = 1
        \right]
        \leq
        n^{-(2/3)k}.
    \end{align*}
    Moreover, note that
    $\bigwedge(f,g) \leq f \land g$.
    As $f \sqcap g = \trim(\cl(\bigwedge(f,g)))$,
    we obtain that
    $(f \land g)(x)=0$
    and
    $(f \apand g)(x) = 1$
    only if
    $
    \cl(\bigwedge(f,g))(x)
    >
    \bigwedge(f,g)(x)
    $.
    Therefore, we also have
    \begin{align*}
        \Pr
        \left[ 
            (f \land g)(\rndn) = 0
            \text{ and }
            (f \apand g)(\rndn) = 1
        \right]
        \leq
        n^{-(2/3)k}.
    \end{align*}
    By the union bound, we conclude:
    \begin{equation*}
        \Pr[C(\rndn) = 0 \text{ and } C^\cala(\rndn) = 1]
        \leq \size(C)\cdot n^{-(2/3)k}.
    \end{equation*}
    This finishes the proof.
\end{proof}

We now prove the lower bound for the clique function.
\begin{theorem}
    \label{thm:clique_lower_bound}
    Let $\delta \in (0,1/3)$
    and $k=n^{1/3-\delta}$.
    The monotone circuit complexity of $\kclq$ is
    $\Omega(n^{\delta^2 k/2})$.
\end{theorem}
\begin{proof}
    Let $C$ be a monotone circuit computing $\kclq$.
    For large $n$,
    we obtain from
    Corollary~\ref{cor:clique_random_graph}
    and Lemmas~\ref{lemma:clique_approx_err}
    and~\ref{lemma:clique_approx_correct}
    \begin{align*}
        5/4
        &\leq
        \Pr[\kclq(\rndy)]
        + 
        \Pr[\kclq(\rndn)]
        \\&
        \leq
        \Pr[C(\rndy) = 1 \text{ and } C^\cala(\rndy) = 0]
        +
        \Pr[C^\cala(\rndy)=1]
        \\&
        \quad +
        \Pr[C(\rndn) = 0 \text{ and } C^\cala(\rndn) = 1]
        +
        \Pr[C^\cala(\rndn)=1]
        \\&
        \leq
        1 + o(1)
        +
        \size(C) \cdot O(n^{-\delta^2 k / 2}).
    \end{align*}
    This implies
    $\size(C) = \Omega(n^{\delta^2 k/2})$.
\end{proof}

\subsection{Proof of Lemma~\ref{lemma:clique_sunflower} 
    (Clique-sunflowers)}
\label{sec:proof_clqsf}

In this section, we give the proof of Lemma~\ref{lemma:clique_sunflower}.
The proof is essentially the same as the one given by Rossman for
Theorem~\ref{thm:approx_sunflower} in~\cite{rossman_kclq}.
We will rely on an inequality due to Janson~\cite{janson1990poisson}
(see also Theorem~2.18 in~\cite{random_graphs}).
\begin{lemma}[Janson's inequality~\cite{janson1990poisson}]
    \label{lemma:janson_inequality}
    Let $\calf$ be a nonempty hypergraph on $[n]$
    and let $\rndW \sseq_p [n]$.
    Define $\mu$ and $\Delta$ in the following way:
    \begin{align*}
        \mu 
        &:=    
        \quad\;\;
        \sum_{F \in \calf}
        \quad\;\;
        \Pr[F \sseq \rndW],
        \\
        \Delta 
        &:= 
        \sum_{\substack{
            F, H \in \calf
            \\
            F \cap H \neq \emptyset
        }}
        \Pr[F \cup H \sseq \rndW].
    \end{align*}
    Then we have
    \begin{equation*}
        \Pr[\forall F \in \calf : F \not\sseq \rndW]
        \leq
        \exp\{-\mu^2/\Delta\}.
    \end{equation*}
\end{lemma}

The following estimates appear in an unpublished note due to
Rossman~\cite{rossman_note_sunflowers},
and a slightly weaker form appears implicitly in~\cite{rossman_kclq}.
We reproduce the proof for completeness.

\begin{lemma}[Lemma~8 of~\cite{rossman_note_sunflowers}]
    Let $s_0(t), s_1(t), \dots$ be the sequence of polynomials defined by
    \begin{equation*}
        s_0(t) := 1
        \quad\text{and}\quad
        s_\ell(t) := 
        t
        \sum_{j=0}^{\ell-1}
        \binom{\ell}{j}
        s_j(t).
    \end{equation*}
    For all $t > 0$,
    we have
    $
        s_\ell(t)
        \leq
        \ell!(t+1/2)^\ell.
    $
\end{lemma}
\begin{proof}
    We first prove by induction on $\ell$ that
    $
        s_\ell(t)
        \leq
        \ell!(\log(1/t+1))^{-\ell}
    $,
    as follows:
    \begin{align*}
        s_\ell(t)
        =
        t \sum_{j=0}^{\ell-1}
        \binom{\ell}{j}
        s_j(t)
        &\leq
        t \sum_{j=0}^{\ell-1}
        \binom{\ell}{j}
        j!(\log(1/t+1))^{-j}
        \\&=
        t
        \ell!(\log(1/t+1))^{-\ell}
        \sum_{j=0}^{\ell-1}
        \frac{(\log(1/t+1))^{\ell-j}}{(\ell-j)!}
        \\&\leq
        t
        \ell!(\log(1/t+1))^{-\ell}
        \left( 
            -1+
        \sum_{j=0}^{\infty}
        \frac{(\log(1/t+1))^{j}}{j!}
        \right)
        \\&=
        t\ell!(\log(1/t+1))^{-\ell}
        (-1+\exp(\log(1/t+1)))
        \\&=
        \ell!(\log(1/t+1))^{-\ell}.
    \end{align*}
    To conclude the proof, we apply the 
    inequality
    $1/\log(1/t+1) < t+1/2$
    for all $t > 0$.
\end{proof}

We will also need the following auxiliary definition.

\begin{definition}
    \label{def:pq_clq_sf}
    Let $\eps,p,q \in (0,1)$.
    Let $\rndU_{n,q} \subseteq [n]$ be a $q$-random subset of $[n]$
    independent of $\rndG_{n,p}$.
    Let $\cals$ be a family of subsets of $[n]$
    and let 
    $B := \bigcap \cals$.
    The family $\cals$ is called a
    $(p, q, \eps)$-\emph{clique-sunflower}
    if
    \[
        \Pr
        \left[
            \exists A \in \cals: 
            \clq{A} \subseteq \rndG_{n,p} \cup \clq{B} 
                \text{ \rm and } 
            A \subseteq \rndU_{n,q} \cup B
        \right] 
        > 1-\eps.
    \]
    The set ${B}$ is called \defx{core}.
\end{definition}

Clearly, a $(p,1,\eps)$-clique sunflower is a
$(p,\eps)$-clique sunflower.
By taking $q=1$ in the following lemma,
and observing that
$s_\ell(\log(1/\eps)) \leq \log(1/\eps)+1/2 \leq 2\log(1/\eps)$ for
$\eps \leq e^{-1/2}$,
we obtain
Lemma~\ref{lemma:clique_sunflower}.

\begin{lemma}
    For all $\ell \in \{1,\dots,n\}$ and $S \subseteq
    \binom{[n]}{\ell}$, if 
    $\abs{\cals} >
    s_\ell(\log(1/\eps)) \cdot (1/q)^\ell(1/p)^{\binom{\ell}{2}}$,
    then 
    $\cals$ contains a $(p,q,\eps)$-clique sunflower.
\end{lemma}

\begin{proof}
    By induction on $\ell$. In the base case $\ell=1$, we have
    by independence that
    \[
        \Pr[ \forall A \in \cals: K_A \nsubseteq \mb G_{n,p} \text{ or } A \nsubseteq \mb U_{n,q}  ]
        &= 
        \Pr[ \forall A \in \cals: A \nsubseteq \mb U_{n,q} ]\\
        &=
        \prod_{A \in S} 
        \Pr[ A \nsubseteq \mb U_{n,q} ]\\
        &=
        (1-q)^{|S|}
         <  
        (1-q)^{\ln(1/\eps)/q}
         \le 
        e^{-\ln(1/\eps)}
         = 
        \eps.
    \]
    Thus $\cals$ is itself a $(p,q,\eps)$-clique sunflower.

    Let now $\ell \ge 2$ and assume that the claim holds for
    $t \in \set{1,\dots,\ell-1}$.
    For convenience, let 
    $$c_j := s_j(\log(1/\eps)),$$ for every 
    $j \in \set{0,1,\dots,\ell-1}$.

    \textit{Case 1.}
    There
    exists $j \in \{1,\dots,\ell-1\}$ and $B \in
    \binom{[n]}{j}$ such that 
    $$|\{A \in \cals : B \subseteq A\}| \ge 
    c_{\ell-j}
    (1/qp^j)^{\ell-j}(1/p)^{\binom{\ell-j}{2}}.$$
    Let $\calt = \{A \setminus B : A \in \cals \text{ such that } B
\subseteq A\} \subseteq \binom{[n]}{\ell-j}$.  
By the
induction hypothesis, 
there exists 
a $(p,qp^j,\eps)$-clique sunflower $\calt' \sseq \calt$
with core a $D$ satisfying
$D \in \binom{[n] \setminus B}{<\ell -j}$.
We will now show that
$\cals' := \set{B \cup C : C \in \calt'} \sseq \cals$
is a $(p,q,\eps)$-clique sunflower contained in $\cals$
with core ${B \cup D}$.
We have
\begin{alignat*}{2}
    \Pr[
        &
        \forall {A} \in \cals':
        \clq{A} \nsubseteq &&\rndG_{n,p} \cup K_{B \cup D} \text{ or } A \nsubseteq \rndU_{n,q} \cup B \cup D
    ]\\
    &=
    \Pr[ 
        \forall C \in \calt':
        &&K_{B \cup C} \nsubseteq \rndG_{n,p} \cup K_{B \cup D} \text{ or } B \cup C \nsubseteq \rndU_{n,q} \cup B \cup D
    ]\\
    &=
    \Pr[ 
        \forall C \in \calt':
        &&K_{B \cup C} \nsubseteq \rndG_{n,p} \cup K_{B \cup D} \text{ or } C \nsubseteq \rndU_{n,q} \cup D
    ]\\
    &=
    \Pr[
        \forall C \in \calt':
        &&K_C \nsubseteq \rndG_{n,p} \cup K_D \text{ or }
        \\
        & \ &&
        C \nsubseteq \big\{v \in \rndU_{n,q} : \{v,w\} \in E(\rndG_{n,p}) \text{ for all } w \in B\big\} \cup D
    ]\\
    &\le
    \Pr[
        \forall C \in \calt':
        &&K_C \nsubseteq \rndG_{n,p} \cup K_D \text{ or } C \nsubseteq \rndU_{n,qp^j} \cup D
    ]\\
    &<
    \eps.&&
\end{alignat*}
Therefore, $\cals'$ is a $(p,q,\eps)$-clique sunflower contained in
$\cals$.

\textit{Case 2.}
For all $j \in \{1,\dots,\ell-1\}$ and
$B \in \binom{[n]}{j}$, we have $$|\{A \in \cals : B \subseteq
A\}| \le
c_{\ell-j}(1/qp^j)^{\ell-j}(1/p)^{\binom{\ell-j}{2}}.$$
In this case, we show that the bound of the lemma holds with $B = \emptyset$.
Let
\[
    \mu &\defeq \abs{\cals} q^\ell p^{\binom\ell 2} > c_\ell,
    \\
    \ovl{\Delta} &\defeq 
    \sum_{j=1}^{\ell-1} \sum_{(A,A') \in \cals^2 \,:\, |A \cap A'| = j}
    q^{2\ell-j} p^{2\binom{\ell}{2} - \binom{j}{2}}.
\]
Note that
$\ovl{\Delta}$ excludes $j = \ell$ from the sum,
which corresponds to pairs $(A,A')$ such that $A = A'$,
in which case the summand becomes $\mu$.
In other words,
the number $\Delta$ of Janson's
inequality~(Lemma~\ref{lemma:janson_inequality})
satisfies
$\Delta = \mu + \ovl{\Delta}$.
Janson's Inequality 
now
gives the following bound:
\begin{equation}\label{eq:janson}
    \Pr[\forall A \in \cals:
        K_A \nsubseteq \mb G_{n,p} \text{ or } 
    A \nsubseteq \mb U_{n,q} ]
    \le
    \exp\left(-
        \frac{\mu^2}{\mu+\ovl{\Delta}}
    \right).
\end{equation}
We bound $\ovl{\Delta}$ as follows:
    \[
        \ovl{\Delta} &\le
        \sum_{j=1}^{\ell-1} q^{2\ell-j} p^{2\binom{\ell}{2} - \binom{j}{2}} \sum_{B \in \binom{[n]}{j}}
        |\{A \in S : B \subseteq A\}|^2 \\
        &\le
        \sum_{j=1}^{\ell-1} q^{2\ell-j} p^{2\binom{\ell}{2} - \binom{j}{2}} \sum_{B \in \binom{[n]}{j}}
        |\{A \in S : B \subseteq A\}| \cdot c_{\ell-j}(1/q)^{\ell-j}(1/p)^{\binom{\ell-j}{2}}\\
        &
        \leq
        q^{\ell}p^{\binom{\ell}{2}} 
        \sum_{j=1}^{\ell-1} c_{\ell-j} \sum_{B \in \binom{[n]}{j}}
        |\{A \in S : B \subseteq A\}|\\
        &=
        q^{\ell}p^{\binom{\ell}{2}}
        \sum_{j=1}^{\ell-1} c_{\ell-j}  
        \sum_{A \in S} \sum_{B \in \binom{A}{j}}
        1\\
        &=
        |S| q^{\ell}p^{\binom{\ell}{2}}
        \sum_{j=1}^{\ell-1}
        \binom{\ell}{j} c_{\ell-j} \\
        &=
        \mu \sum_{j=1}^{\ell-1} \binom{\ell}{j}  c_j 
        =
        \mu \sum_{j=0}^{\ell-1} \binom{\ell}{j}  c_j - \mu.
    \]
Therefore, 
    \[
        \frac{\mu^2}{\mu + \ovl{\Delta}}
        \ge
        \frac{\mu}{\sum_{j=0}^{\ell-1} \binom{\ell}{j}  c_j }
        =
        \frac{\mu}{c_\ell/(\log(1/\eps))}
        >
        \log(1/\eps).
    \]
Finally, from (\ref{eq:janson}) we get 
    \[
    \Pr[\forall A \in \cals:
        K_A \nsubseteq \mb G_{n,p} \text{ or } 
    A \nsubseteq \mb U_{n,q} ]
    \le
    \exp\left(-
        \frac{\mu^2}{\mu+\ovl{\Delta}}
    \right)
    < \eps.
    \]
    Therefore, the family $\cals$ is a $(p,q,\eps)$-clique sunflower with an empty
    core.
\end{proof}

\section{Monotone arithmetic circuits}
\label{sec:arithmetic}

In this section, we give a short and simple proof of a truly exponential
($\exp(\Omega(n))$) lower bound for real monotone 
arithmetic
circuits
computing a multilinear $n$ variate polynomial. 
Real monotone arithmetic circuits are arithmetic circuits
over the reals
that use only positive numbers as coefficients.
As we shall see, the
lower bound argument holds for a general family of multilinear
polynomials constructed in a very natural way from error correcting
codes, and the similarities to the hard function used by Harnik and Raz
in the Boolean setting is quite evident
(see Section~\ref{sec:hr_the_function}). In particular, our lower bound
just depends on the rate and relative distance of the underlying code. We
note that exponential lower bounds for monotone 
arithmetic
circuits are not
new, and have been known since the 80's with various quantitative bounds.
More precisely, Jerrum and Snir proved an $\exp(\Omega(\sqrt{n}))$ lower
bound for an $n$ variate polynomial in \cite{JS82}. This bound was
subsequently improved to a lower bound of $\exp(\Omega(n))$ by Raz and
Yehudayoff in \cite{RY11}, via an extremely clever argument, which relied
on deep and beautiful results on character sums over finite fields. A
similar lower bound of $\exp(\Omega(n))$ was shown by Srinivasan \cite{S19}
using more elementary techniques building on a work of Yehudayoff
\cite{Y19}.  In a recent personal communication Igor Sergeev pointed out to
us that truly exponential lower bounds for monotone arithmetic circuits had
also been proved in the 1980's in the erstwhile Soviet Union by several
authors, including the works of Kasim-Zade, Kuznetsov and Gashkov. We refer
the reader to \cite{GSergeev2012} for a detailed discussion on this line of
work. 

We show a similar lower bound of $\exp(\Omega(n))$ via a simple and short
argument, which holds in a somewhat general setting. Our contribution is
just the simplicity, the (lack of) length of the argument and the
observation that it holds for families of polynomials that can be
constructed from any sufficiently \emph{good} error correcting codes.

\begin{definition}
    [Monotone, multilinear, homogeneous]
    \label{def:arithmetic_glossary}
    A real polynomial is said to \emph{monotone}
    if all of its coefficients are positive.
    A real 
    arithmetic 
    circuit
    is said to be \emph{monotone} if it uses
    only positive numbers as coefficients.
    A polynomial $P$ is said to be \emph{multilinear} if the degree of each
    variable of $P$ is at most $1$ in all of the monomials of $P$. 
    A polynomial $P$ is said to be \emph{homogeneous} if
    all the monomials of $P$ have the same degree.
    An arithmetic circuit $C$ is said to be to \emph{homogeneous}
    (\emph{multilinear}) if the polynomial
    computed in each of the gates of $C$ is homogeneous (multilinear).
\end{definition}

\begin{definition}[From sets of vectors to polynomials]\label{def:char poly}
    Let ${C} \subseteq \F_q^n$ be an arbitrary subset of $\F_q^n$. Then,
    the polynomial $P_{C}$ is a multilinear homogeneous
    polynomialof
    degree $n$ on $qn$ variables 
    $\{x_{i, j} : i \in [q], j \in [n]\}$ 
    and
    is defined as follows:
    \[
        P_C = \sum_{c \in C} \prod_{j \in [n]} 
        x_{c(j), j} \, .
    \]
    Here, $c(j)$ is the $j^{th}$ coordinate of $c$ which is an element of $\F_q$, which we bijectively identify with the set $[q]$. 
\end{definition}

Here, we will be interested in the polynomial $P_C$ when the set $C$ is a \emph{good} code, i.e it has high rate and high relative distance. The following observation summarizes the properties of $P_C$ and relations between the properties of $C$ and $P_C$. 

\begin{observation}[Codes vs Polynomials]\label{obs:props of poly}
    Let $C$ be any subset of $\F_q^n$ and let $P_C$ be the polynomial as defined in Definition~\ref{def:char poly}. Then, the following statements are true: 
    \begin{enumerate}[$\bullet$]
        \item $P_C$ is a multilinear homogeneous polynomial of degree equal to $n$ with every coefficient being either $0$~or~$1$.
        \item The number of monomials with non-zero coefficients in $P_C$ is equal to the cardinality of $C$.
        \item If any two distinct vectors in $C$ agree on at most $k$ coordinates (i.e. $C$ is a code of distance $n-k$), then the intersection of the support of any two monomials with non-zero coefficients in $P_C$ has size at most~$k$.  
    \end{enumerate}
\end{observation} 
The observation immediately follows from Definition~\ref{def:char poly}. We
note that we will work with monotone 
arithmetic
circuits here, and hence
will interpret the polynomial $P_C$ as a polynomial over the field of real
numbers. 

We now prove the following theorem, which essentially shows that for
every code $C$ with sufficiently good distance, any monotone 
arithmetic
circuit computing $P_C$ must essentially compute it by computing each of
its monomials separately, and taking their 
sum. 
\begin{theorem}\label{thm:monotone alg lower bound}
    If any two distinct vectors in $C$ agree on at most $n/3-1$ locations, then any monotone 
    arithmetic
    circuit for $P_C$ has size at least $|C|$. 
\end{theorem}

The proof of this theorem crucially uses the following well known structural lemma about 
arithmetic
circuits. This lemma also plays a crucial role in the other proofs of exponential lower bounds for monotone 
arithmetic
circuits (e.g. \cite{JS82,RY11,Y19,S19}). 

\begin{lemma}[See Lemma 3.3 in~\cite{RY11}]\label{lem:decomposing a size s circuit}
    Let $Q$ be a homogeneous multilinear polynomial of degree $d$ computable by a homogeneous 
    arithmetic
    circuit of size $s$. Then, there are homogeneous polynomials $g_0, g_1, g_2, \ldots, g_s, h_0, h_1, h_2, \ldots, h_s$ of degree at least $d/3$ and at most $2d/3-1$ such that 
    \[
        Q = \sum_{i = 0}^{s} g_i\cdot h_i \, .
    \]
    Moreover, if the circuit for $Q$ is monotone, then each $g_i$ and $h_i$ is multilinear, variable disjoint and each one their non-zero coefficients is a positive real number.
\end{lemma} 
We now use this lemma to prove Theorem~\ref{thm:monotone alg lower bound}.
\begin{proof}[Proof of Theorem~\ref{thm:monotone alg lower bound}]
    Let $B$ be a monotone 
    arithmetic
    circuit for $P_C$ of size $s$. We know from Observation~\ref{obs:props of poly} that $P_C$ is a multilinear homogeneous polynomial of degree equal to $n$. This along with the monotonicity of $B$ implies that $B$ must be homogeneous and multilinear since there can be no cancellations in $B$. Thus, from (the moreover part of) Lemma~\ref{lem:decomposing a size s circuit} we know that $P_C$ has a monotone decomposition of the form
    \[
        P_C = \sum_{i = 0}^s g_i\cdot h_i \, ,
    \] 
    where, each $g_i$ and $h_i$ is multilinear, homogeneous with degree between $n/3$ and $2n/3-1$,  $g_i$ and $h_i$ are variable disjoint. We now make the following claim.
    \begin{claim}\label{clm:single monomial}
        Each $g_i$ and $h_i$ has at most one non-zero monomial. 
    \end{claim}
    We first observe that the claim immediately implies theorem~\ref{thm:monotone alg lower bound}: since every $g_i$ and $h_i$ has at most one non-zero monomial, their product $g_ih_i$ is just a monomial. Thus, the number of summands $s$ needed in the decomposition above must be equal to the number of monomials in $P_C$, which is equal to $|C|$ from the second item in Observation~\ref{obs:props of poly}. 
\end{proof}
We now prove the Claim. 
\begin{proof}[Proof of Claim]
    The proof of the claim will be via contradiction. To this end, let us assume that there is an $i \in \{0, 1, 2, \ldots, s\}$ such that $g_i$ has at least two distinct monomials with non-zero coefficients and let  $\alpha$ and $\beta$ be two of these monomials. Let $\gamma$ be a monomial with  non-zero coefficient in $h_i$ . Since $h_i$ is homogeneous with degree between $n/3$ and $2n/3-1$, we know that the degree of $\gamma$ is at least $n/3$. Since we are in the monotone setting, we also know that each non-zero coefficient in any of the $g_j$ and $h_j$ is a positive real number. Thus, the monomials $\alpha\cdot \gamma$ and $\beta\cdot \gamma$ which have non-zero coefficients in the product $g_i\cdot h_i$ must have non-zero coefficient in $P_C$ as well (since a monomial once computed cannot be cancelled out). But, the supports of $\alpha\gamma$ and $\beta\gamma$ overlap on $\gamma$ which has degree at least $n/3$. This contradicts the fact that no two distinct monomials with non-zero coefficients in $P_C$ share a sub-monomial of degree at least $n/3$ from the distance of $C$ and the third item in Observation~\ref{obs:props of poly}. 
\end{proof} 
Theorem~\ref{thm:monotone alg lower bound} when instantiated with an appropriate choice of the code $C$, immediately implies an exponential lower bound on the size of monotone 
arithmetic
circuits computing the polynomial $P_C$. Observe that the total number of variables in $P_C$ is $N = qn$ and therefore, for the lower bound for $P_C$ to be of the form $\exp(\Omega(N))$, we would require $q$, the underlying field size to be a constant. In other words, for any code of relative distance at least $2/3$ over a constant size alphabet which has exponentially many code words, we have a truly exponential lower bound. 

The following theorem of Garcia and Stichtenoth~\cite{GS95} implies an explicit construction of such codes. The statement  below is a restatement of their result by Cohen et al.\cite{CHS18}. 
\begin{theorem}[\cite{GS95} and \cite{St09}]
Let $p$ be a prime number and let $m\in \N$ be even. Then, for every $0 <\rho  < 1$ and a large enough integer $n$, there exists an explicit rate $\rho$ linear error correcting block code $C: \F_{p^m}^n \to \F_{p^m}^{n/\rho}$ with distance 
\[
\delta \geq 1- \rho - \frac{1}{p^{m/2} - 1} \, .
\]
\end{theorem}
The theorem has the following immediate corollary. 
\begin{corollary}\label{cor:good codes exist}
    For every large enough constant $q$ which is an even power of a  prime, and for all large enough $n$,  there exist explicit construction of codes $C \subseteq \F_q^n$ which have relative distance at least $2/3$ and $|C| \geq \exp(\Omega(n))$.
\end{corollary}
By an explicit construction here, we mean that given a vector $v$ of length $n$ over $\F_q$, we can decide in deterministic polynomial time if $v \in C$. In the 
arithmetic
complexity literature, a polynomial $P$ is said to be explicit, if given the exponent vector of a  monomial, its coefficient in $P$ can be computed in deterministic polynomial time. Thus, if a code $C$ is explicit, then the corresponding polynomial $P_C$ is also explicit in the sense described above. Therefore, we have the following corollary of Corollary~\ref{cor:good codes exist} and Theorem~\ref{thm:monotone alg lower bound}.

\begin{corollary}
    There exists an explicit family $\{P_n\}$ of homogeneous multilinear polynomials such that for every large enough $n$, any monotone 
    arithmetic
    circuit computing the $n$ variate polynomial $P_n$ has size at least $\exp(\Omega(n))$.  
\end{corollary}

\section{Further directions}

In this paper, we obtained the first monotone circuit lower bound of the form
$\exp(\Omega(n^{1/2}/\log n))$
for an explicit $n$-bit monotone Boolean function.
It's natural to ask if we can do better.
Ideally, we would like to achieve a truly exponential bound
for Boolean monotone circuits, like the one achieved for
arithmetic monotone circuits in Section~\ref{sec:arithmetic}.
However, as discussed in Section~\ref{sec:hr_discussion},
the $\sqrt{n}$ exponent seems to be at the limit
of what current techniques can achieve.

An important open-ended direction is to
develop sharper techniques for proving monotone circuit lower bounds.
Sticking to the approximation method,
it is not yet known
whether there exists another ``sunflower-type'' notion
which still allows for good approximation bounds
and yet
admits significantly better bounds than what is possible
for robust sunflowers.

One approach can be to try to weaken the requirement of the core,
and ask only that the core 
of a ``sunflower-type'' set system $\calf$
is properly contained in one of the elements of $\calf$.
A weaker notion of robust sunflowers with this weakened 
core
could still be used succesfully in the proof of the lower bound of
Section~\ref{sec:harnik_raz},
but it's not yet clear
whether this weaker notion admits stronger bounds or not.

Moreover, perhaps developing specialised sunflowers for specific functions, such as
done for $\kclq$ in Section~\ref{sec:clique}, could help here.
One could also consider distributions which are not $p$-biased,
as perhaps better bounds are possible in different regimes.

Finally, as noted before,
our proof of the clique-sunflower lemma follows the approach
of Rossman in~\cite{rossman_kclq}.
We expect that a proof along the lines of 
the work of
Alweiss, Lovett, Wu and Zhang~\cite{alweiss2019improved}
and
Rao~\cite{rao2020}
should 
give us an even better bound
on the size of set systems without clique-sunflowers, 
removing the $\ell !$ factor. 
This would extend our $n^{\Omega(\delta^2 k)}$ lower bound to $k \le
n^{1/2-\delta}$.

\section*{Acknowledgements}
We are grateful to Stasys Junka for bringing the lower bound of Andreev
\cite{andreev1987method} to our attention and to the anonymous referees of
LATIN 2020 for numerous helpful suggestions. We also thank Igor Sergeev for
bringing \cite{GSergeev2012} and the references therein to our attention
which show that truly exponential lower bounds for monotone arithmetic
circuits had already been proved in the 1980s. 
Finally, we thank the anonymous reviewers of Algorithmica
for careful proofreading and many helpful suggestions and comments.

Bruno Pasqualotto Cavalar
was supported by 
S\~ao Paulo Research Foundation (FAPESP),
grants \#2018/22257-7 and \#2018/05557-7,
and he acknowledges
CAPES (PROEX)
for partial support of this work.
A part of this work was done during a research internship of 
Bruno Pasqualotto Cavalar
and a postdoctoral stay of 
Mrinal Kumar
at the
University of Toronto.
Benjamin Rossman
was supported by NSERC and Sloan Research
Fellowship. 

This version of the article has been accepted for publication after peer review,
but is not the Version of Record and does not reflect post-acceptance improvements, or any
corrections. The Version of Record is available online at:
\url{https://doi.org/10.1007/s00453-022-01000-3}.

\bibliographystyle{plain}%
\bibliography{bibliography}

\appendix

\section{Proof of Theorem~\ref{thm:improved_sunflower}}
\label{sec:sf_proof}

We say that 
a
family $\calf$ of sets is 
\emph{$r$-spread}
if
there are most 
$\card{\calf}/r^{\card{T}}$
sets in $\calf$
containing any given non-empty set $T$.
The following theorem is a $p$-biased variant of the main technical
lemma of Rao~\cite{rao2020}. A full proof is given in the appendix
of~\cite{bcw21}.

\begin{theorem}[Theorem 3 of~\cite{bcw21}]
    \label{thm:spread}
    There exists a constant $B > 0$ such that the following holds
    for all $p, \eps \in (0,1/2]$ and all positive integers $\ell$.
    Let $r = B \log(\ell/\eps)/p$.
    Let 
    $\calf$ be a $r$-spread $\ell$-uniform family
    of subsets of $[n]$
    such that
    $\abs{\calf} \geq 
    r^{\ell}$.
    Then 
    $\Pr_{\rndW \sseq_p [n]} [\exists F \in \calf : F \sseq \rndW] >
    1-\eps$.
\end{theorem}

We now combine Theorem~\ref{thm:spread} with the main argument
of the proof of Theorem~4.4 of~\cite{rossman_kclq}
to finish the proof of Theorem~\ref{thm:improved_sunflower}.

\begin{proof}[Proof of Theorem~\ref{thm:improved_sunflower}]
    The proof is by induction on $\ell$. When $\ell=1$, we have
    \begin{equation*}
        \Pr[\forall F \in \calf : F \not\sseq \rndW]
        =
        (1-p)^{\abs{\calf}}
        \leq
        e^{-p \abs{\calf}}
        <
        \eps.
    \end{equation*}
    Therefore, $\calf$ itself is a $(p,\eps)$-robust sunflower.
    We now suppose $\ell > 1$ and that the result holds for every
    $t \in [\ell-1]$.
    For a set $T \sseq [n]$, let 
    $\calf_T = \set{F \sm T : F \in \calf, T \sseq F}$.
    Let $r = B \log(\ell/\eps)/p$,
    where $B$ is the constant of Theorem~\ref{thm:spread}.

    \textit{Case 1.}
    The family $\calf$ is not $r$-spread.
    By definition,
    there exists a nonempty set $T \sseq [n]$
    such that
    $\abs{\calf_T} > \card{\calf}/r^{\card{T}} \geq r^{\ell-\abs{T}}$.
    By induction,
    the family $\calf_T$
    contains a $(p,\eps)$-robust sunflower $\calf'$.
    It is easy to see that
    $\set{F \cup T : F \in \calf'}$ is a 
    $(p,\eps)$-robust sunflower contained in $\calf$.

    \textit{Case 2.}
    The family $\calf$ is $r$-spread.
    Therefore, from Theorem~\ref{thm:spread},
    it follows that $\calf$ is itself a $(p,\eps)$-robust sunflower.
\end{proof}

\end{document}